\documentclass[twocolumn,showpacs,preprintnumbers,amsmath,amssymb,a4paper,superscriptaddress, nofootinbib,aps,pra,10pt]{revtex4-1}

\usepackage[cp1251]{inputenc}  \usepackage[english]{babel}    

\usepackage{float}
\usepackage{amsmath}
\usepackage{graphicx}
\usepackage{dcolumn}
\usepackage{bm}

\usepackage{epstopdf}

\usepackage{latexsym}
\usepackage{amssymb}
\usepackage{eqnarray} 

\usepackage{mathrsfs}
\usepackage[percent]{overpic}

\begin{document}

\selectlanguage{english} 

\title{
Aspects of arbitrarily oriented dipoles scattering in plane: short-range interaction influence
}
\author{Eugene A. Koval}
\email[]{e-cov@yandex.ru}
\affiliation{Bogoliubov Laboratory of Theoretical Physics, Joint Institute for Nuclear Research, Dubna, Moscow Region 141980, Russian Federation}

\author{Oksana A. Koval}
\email[]{kov.oksana20@gmail.com}
\affiliation{A.M. Obukhov Institute of Atmospheric Physics, Russian Academy of Sciences, Moscow 119017, Russian Federation}

\date{\today}

\begin{abstract}\label{txt:abstract}
The impact of the short-range interaction on the resonances occurrence in the anisotropic dipolar scattering in a plane was numerically investigated \textit{for the arbitrarily oriented dipoles} and for a wide range of collision energies. 
We revealed the strong dependence of the cross section of the 2D dipolar scattering on the radius of short-range interaction, which is modeled by a hard wall potential and by the more realistic Lennard-Jones potential, and on the mutual orientations of the dipoles. 
We defined the critical (magic) tilt angle of one of the dipoles, depending on the direction of the second dipole for arbitrarily oriented dipoles. It was found that resonances arise only when this angle is exceeded. 
In contrast to the 3D case, the energy dependencies of the boson (fermion) 2D scattering cross section grows (is reduced) with an energy decrease in the absence of the resonances. 
We showed that the mutual orientation of dipoles strongly impacts the form of the energy dependencies, which begin to oscillate with the tilt angle increase, unlike the 3D dipolar scattering. The angular distributions of the differential cross section in the 2D dipolar scattering of both bosons and fermions are highly anisotropic at non-resonant points. The results of the accurate numerical calculations of the cross section agree well with the results obtained within the Born and eikonal approximations. 
\end{abstract}

\maketitle

\section{Introduction}
\label{sec:Intro}

The three-dimensional (3D) dipolar scattering problem has been studied extensively (e.g.~\cite{Baranov_2008, Lahaye_2009, tang2018tuning, Cavagnero_2009, zhang2014effect,chomaz2019long, tanzi2019observation,frisch2014quantum, maier2015emergence, yang2017classical}), whereas the two-dimensional (2D) dipolar scattering was separated to the stand-alone topic, that has been actively developed and has attracted increasing interest in a number of theoretical~\cite{bohn2017coldREVIEW,de2011controlling,micheli2007cold,Baranov_2008, Bloch_2008, micheli2006toolbox, baranov2012condensed, Lahaye_2009, Ye_2018, Gorecki_2017, aikawa2014reaching} and experimental~\cite{bohn2017coldREVIEW,de2011controlling, Ye_2018, Gorecki_2017, aikawa2014reaching} studies recently.

Dipolar gases are more stable in a quasi-two-dimensional geometry, in contrast with a 3D case, due to the absence of the "head-to-tail" instability~\cite{bohn2017coldREVIEW,Pfau_2008,Ni_2010}. 
Optical lattices with 2D geometries are prospective candidates for a dipolar gas stabilization, trapping, and dynamics controlling because the dipole-dipole interaction (DDI) is isotropic and repulsive in the case of dipole moments polarized along the frozen direction, whereas tilting of the polarization axis leads to a controllable anisotropy of the interaction~\cite{bohn2017coldREVIEW,de2011controlling,Lahaye_2009}. 
Molecules collisions in one layer of a pancake-shaped trap are modelled by a 2D dynamics of the molecules~\cite{de2011controlling,Ticknor_2009,Ticknor_2011,Ticknor_Bohn_2011,Volosniev_2011,Rosenkranz_2011,Koval_2014}.
The investigations of a dipolar diatomic molecules interaction in a plane are highly relevant due to prospects for one of the possible realization of a qubit and application to the quantum computing schemes~\cite{DeMille_2002,ni2018dipolar}. 

As noted in Refs.~\cite{Ye_2018,Ticknor_2009}, the ongoing experiments call for a deeper understanding of the short-range physics influence on dipolar scattering in a plane. From the experimental point of view, the dipoles' short-range interaction control is possible via external fields and Feshbach resonances mechanisms~\cite{Pfau_2008,Lahaye_2009}.
There is a number of significant differences of a 2D dipolar scattering in contrast to the 3D scattering, e.g. an $s-$wave divergence in the low-energy limit and an existence of a weakly bound state for any attractive potential~\cite{Simon_1976}. 
The known results, describing shape resonances in dependencies of the cross section on the short-range interaction (SRI) radius (cut-off radius) for 3D space~\cite{Cavagnero_2009}, are inapplicable to the 2D dipolar scattering.
The dependencies of the 2D dipolar scattering cross section for the fixed SRI radius were studied in Refs.~\cite{Ticknor_2011,Koval_2014}.
The \textit{two-dimensional} dipolar scattering cross section dependencies on the SRI radius for the arbitrary orientation of the dipole moments have not been studied. This research fills this gap.

We determined, that for the scattering in a plane cross sections for fermions are substantially smaller, than those for bosons at low energies of collisions, like in the low-energy 3D dipolar scattering~\cite{Cavagnero_2009}. Cross sections of a 2D dipolar scattering of fermions at high energies are comparable with the cross sections of bosons. 
We demonstrated the resonances' occurrence and an increase of their number at the decrease of SRI radius below the ``threshold'' values obtained in this paper.
In the absence of the resonances the energy dependencies of the boson (fermion) dipolar scattering cross section grows (is reduced) with energy decrease in 2D case, in contrast to the 3D case, where it has the form of a plateau for both bosons and fermions~\cite{Cavagnero_2009}. 
We showed that the mutual orientation of dipoles strongly impacts the form of the resonant and non-resonant energy dependencies, which begin to oscillate with the tilt angle increase, unlike the 3D dipolar scattering~\cite{Cavagnero_2009,bohn2009quasi}, where there is no such oscillations.

In contrast to a large amount of the papers, devoted to the scattering of \textit{aligned} dipoles~\cite{ronen2006dipolar,Ticknor_2009,Ticknor_2011,Kanjilal_2006,Cavagnero_2009,hanna2012resonant,Lahaye_2009} we consider the scattering of \textit{arbitrarily oriented} dipoles in a plane. Anisotropic collisions of two dipolar Bose-Einstein condensates~\cite{burdick2016anisotropic} or collisions of slow polar molecules prepared in a ``cryofuge''~\cite{wu2017cryofuge}, make the two-body differential scattering cross section detectable. The collided dipoles will have different orientations of dipole moments if the initial samples of dipolar gases are prepared under differently directed external electric fields. In this paper, the strong changes of the differential cross section angular distributions in resonant and non-resonant points were revealed for the scattering of the arbitrarily oriented dipoles in a plane, not only for fermions but for bosons as well. So, this paper is highly relevant for an interpretation of such experiments.

The dipoles interaction potential at short-range is approximated here by two types of potentials: by a hard wall with the radius $\rho_{SR}$~\cite{ronen2006dipolar,Ticknor_2011,Kanjilal_2006,Cavagnero_2009,Ticknor_2009,Koval_2014} and with a more realistic Lennard-Jones potential~\cite{hanna2012resonant, Hutson2018near}, while at long-range it is modelled by the DDI~\cite{Greene_2015}.

The description for the 2D dipolar scattering problem and the improved numerical algorithm (with the better convergence) for its solution are briefly described in Sec.~\ref{sec:2DDipoleScattering}. The analysis of the numerically obtained results and their comparison with the results within the Born and eikonal approximations are presented in Sec.~\ref{sec:Results}. Sec.~\ref{sec:Conclusions} contains the conclusion remarks.

\section{Quantum scattering of arbitrarily oriented dipoles in plane}
\label{sec:2DDipoleScattering}

The 2D Shr\"{o}dinger equation for describing quantum dipolar scattering in a plane on anisotropic potential $U(\rho ,\phi )$ in polar coordinates $(\rho ,\phi )$ reads:
\begin{equation}
\label{eq2DShroedingerEquationBasic}
\Bigl[ 
-\frac{\hbar ^2}{2\mu }
\left( 
{\frac{1}{\rho }\frac{\partial }{\partial \rho }\left( {\rho \frac{\partial }{\partial \rho  }} \right)+\frac{1}{\rho ^2}\frac{\partial ^2}{\partial \phi ^2}} \right)+U\left( {\rho ,\phi } 
\right) -E
\Bigr] 
\Psi \left( {\rho ,\phi } \right)=0 \end{equation}
with boundary condition in the asymptotic region ${\rho \to \infty }$:
\begin{equation}
\label{eqBoundaryConditionAtInfinity}
\Psi \left( {\rho ,\phi } \right) \to e^{i {\bm q}\bm{\rho}}
+f\left( {q,\phi ,\phi _q } \right)
\frac{e^{iq\rho }}{\sqrt {-i\rho } }.
\end{equation}
The relative momentum ${\bm q}$ is defined by the collision energy $E$ with $q= \sqrt {2\mu E}/\hbar$ and $\mu $~denotes the reduced mass of the system. The incoming wave direction ${\bm q}/q$ is defined by the $\phi_q$ angle. 

 \begin{figure}[tbp]
 \centerline{\includegraphics[width=0.7\linewidth]{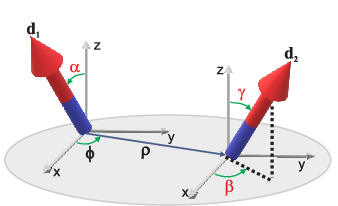}}
 \caption{The scheme of mutual orientation of two arbitrarily oriented dipoles ${\bm d}_1 $ and ${\bm d}_2 $, moving in the $XY$ plane.}
 \label{figSchemeOfTwoDipoles}
 \end{figure}
 
The interaction potential has the form: 
\begin{equation}
 \label{eqPotentialU}
 U(\rho,\phi) = V_{SR}(\rho) + V_{dd}(\rho,\phi),
 \end{equation}
 
The long-range interaction potential of the two arbitrarily oriented dipoles~$V_{dd}(\rho,\phi)$ in a plane reads:
 \begin{equation}
 \label{eqPotentialVDipoleDipole}
 V_{dd}({\bm \rho}, {\bm d}_1, {\bm d}_2) =\frac{1}{\rho
^3}\left(({\bm d}_1 {\bm d}_2 )-3\frac{({\bm d}_1 {\bm \rho
})({\bm d}_2 {\bm \rho })}{\rho ^2}\right),
\end{equation}
where ${\bm d}_i (i=1,2)$ -- dipole moments, and $({\bm d}_i {\bm \rho })/\rho$ denote their projections on the axis, that connect dipole centres of mass. The expression~(\ref{eqPotentialVDipoleDipole}) in polar coordinates is represented as:
 \begin{align}
 \label{eqPotentialVDipoleDipoleInPolarCoordinates}
  V_{dd}\left( {\rho ,\phi; \alpha ,\beta ,\gamma } \right)= \frac{d_1 d_2
 }{\rho ^3}[\sin (\alpha )\sin (\gamma )\cos (\beta ) + \nonumber \\
  +\cos (\alpha )\cos
 (\gamma )
  -3\sin (\alpha )\sin (\gamma )\cos (\phi )  \cos (\phi - \beta)],
 \end{align}
\noindent where angles $\alpha$ and $\gamma$ define the dipole tilt with respect to the $Z$-axis and the angle $\beta$ determines the spatial orientation of the $Zd_1$ and $Zd_2$ planes. The scheme of the mutual orientation of two arbitrarily oriented dipoles is represented in Fig.~\ref{figSchemeOfTwoDipoles}.

In this paper, for an approximation of the repulsive SRI $V_{SR}(\rho)$ we use two different types of potentials: 
the hard wall potential with the width of $\rho_{SR}$ (so that the wave function is equal to zero at $\rho_{SR}$):
\begin{equation}
\label{eqPotentialVShortRange}
V_{SR}(\rho) = \left\{ {{\begin{array}{*{20}c}
  {\infty ,\mbox{ }\rho \leqslant \rho _{SR}} \hfill \\
  {0,\mbox{ }\rho >\rho _{SR}} \hfill \\
 \end{array} }} \right.,
\end{equation}
which has been previously applied by other authors~\cite{Kanjilal_2006,Cavagnero_2009,Ticknor_2011}, and \\
the more realistic Lennard-Jones (LJ) potential~\cite{Hutson2018near, hanna2012resonant} 
\begin{equation}
\label{eqPotentialLJ}
V_{SR}(\rho) = \frac{ C_{12} }{ \rho^{12} } - \frac{C_{6}}{\rho^6}.
\end{equation}
A fixed value of the $C_{6}$ parameter of the Lennard-Jones potential was taken for polar molecules, e.g., $C_{6}= 1.5\mbox{ }10^6$ a.u. for the $^{23}Na^{87}Rb$ polar molecule~\cite{gonzalez2017adimensional, Hutson2018near}. The dipolar length scale $D$ for polar molecules (${D \approx 182554}$ a.u. for $^{23}Na^{87}Rb$) is much larger than the van der Waals length scale ${R_6 = (2 \mu C_6 / \hbar^2)^{1/4}}$ ({${\mu = 100167}$ a.u.}; ${R_6 \approx 740}$ a.u. for $^{23}Na^{87}Rb$)~\cite{Hutson2018near} and the Lennard-Jones potential is effectively used as the short-range repulsion potential for the dipole-dipole interaction potential. Increasing the $C_{12}$ at the fixed $C_{6}$ leads to an increase of the short-range part of the $U(\rho,\phi)$. So we vary $C_{12}$ to reproduce the change of $\rho_{SR}$.

In order to compare the results of two types of potential with each other, we need to relate the SRI radius $\rho_{SR}$ with the Lennard-Jones $C_{6},C_{12}$ parameters.
We define $\rho_{SR}$ for the Lennard-Jones potential as $\rho_{SR} =\min(\rho_0(\phi))$, where $\rho_0(\phi)$ is the positions of the zeros of the potential $U(\rho,\phi)$.
From the physical point of view, $\min(\rho_0(\phi))$~--- the minimal distance, that molecules could reach at low energies of collision.
For the considered polar molecules the term $-C_{6}/\rho^6$ is small compared to $V_{dd}$. 
Thus, $\rho_{SR}$ is found under the condition ${C_{12}/\rho^{12} + V_{dd}(\rho,\beta/2) = 0}$, from which the next relation 
follows:
\begin{equation}
\label{eqRoLJVSRoSR}
\rho_{SR} = \left[ \frac{ C_{12} (E_D D^{3} )^{-1} }{  \sin(\alpha)\sin(\gamma)  \frac{3+ \cos(\beta)}{2}  - \cos(\alpha)\cos(\gamma) } \right]^{\tfrac{1}{9}}.
\end{equation}

For the hard wall potential~(\ref{eqPotentialVShortRange}), the equation~(\ref{eq2DShroedingerEquationBasic}) with potential~(\ref{eqPotentialU}) in the dipole units of length $D$ and energy $E_D$:
\[
D=\mu d^2/\hbar^2, \newline
E_D=\hbar^6/\mu^3 d^4,
\]
could be written in a dimensionless form:
\begin{align}
 \label{eq2DShroedingerEquation}
\left[-\frac{1}{2 }\left( \frac{\partial^2 }{\partial \rho^2 }+ {\frac{1}{\rho
 }\frac{\partial }{\partial \rho }+\frac{1}{\rho ^2}\frac{\partial ^2}{\partial \phi ^2}} \right)
 +U\left( {\rho
 ,\phi } \right) \right]
\Psi=E\Psi
\end{align}
For the $^{23}Na^{87}Rb$ polar molecule the dipole moment {$d=1.35$ a.u.}~\cite{gonzalez2017adimensional}, so {$E_D = 2.996 \times 10^{-16}$ a.u.}. We note, that in case the hard wall potential is chosen as $V_{SR}$, the dipolar scattering properties can be scaled for the parameters of the particular system~\cite{bohn2009quasi}. For the consideration of the Lennard-Jones potential, the dipolar scattering properties depend on the relative strength of the van der Waals and dipolar interactions.  The Lennard-Jones potential conversion to the dipolar units reads:
\begin{equation}
\label{eqPotentialLJinDipolarUnits}
V_{SR}(\rho) = \frac{1}{E_D} \left( \frac{ C_{12} }{ \rho^{12} } - \frac{C_{6}}{\rho^6} \right).
\end{equation}

A scattering differential cross section is defined by the calculated scattering amplitude $f\left( {q,\phi ,\phi _q } \right)$ 
 \begin{equation}
 \label{eqDiffCrossSectionDefinition}
 {d\sigma (q,\phi,\phi_q )} \mathord{\left/ {\vphantom {{d\sigma (q,\Omega )}
 {d\Omega }}} \right. \kern-\nulldelimiterspace} {d\Omega
 }\mbox{=}{\left| {f(q,\phi ,\phi
 _q )} \right|^2},\,
 \end{equation}
where $d\Omega = d\phi_q d\phi$. A total cross section is obtained by averaging over incoming wave directions ($\phi_q$) and integration over scattering angle $\phi$:
 \begin{equation}
 \label{eqFullCrossSectionDefinition}
 \sigma (q)\mbox{=}\frac{1}{2\pi }\int\limits_0^{2\pi }
 \int\limits_0^{2\pi } {\frac{d\sigma }{d\Omega }}
d\phi _q d\phi.
 \end{equation}
 
A transition to scattering of identical bosons (fermions) is done with the symmetrization ${\epsilon = +1}$ (antisymmetrization ${\epsilon = -1}$) of the wave function:
\begin{equation}
 \label{eqAsymptoticsWaveFunctionBosonFermion}
\Psi \left( {\rho ,\phi } \right)\to e^{i {\bm q}\bm{\rho}  } + \epsilon \, e^{-i {\bm q}\bm{\rho} } + f(\phi) \frac{e^{iq\rho }}{\sqrt {-i\rho } }
\end{equation}
as well as the differential cross section:
\begin{equation}
\label{eqDiffCrossSectionDefinitionBosonFermion}
d\sigma (\phi)/{d\Omega }=\left| {f(\phi,\phi_q)} \right|^2 =
\left| f\left( {\phi} \right) + \epsilon f\left( {|180^{\circ} - \phi|} \right) \right|^2.
\end{equation}

The definitions~(\ref{eqDiffCrossSectionDefinition}) and~(\ref{eqDiffCrossSectionDefinitionBosonFermion}) show, that it leads to an increase for bosons and a decrease for fermions of the differential cross section with respect to the case of distinguishable particles for some scattering angles, e.g., for $\phi + \phi_q = 90^{\circ}$: 
\begin{align*}
\frac{d\sigma_{B}(\phi + \phi_q)}{d\Omega } & = 4\frac{d\sigma(\phi + \phi_q)}{d\Omega } \mbox{ (for bosons)},
\\
\frac{d\sigma_{F}(\phi + \phi_q)}{d\Omega } & = 0 \mbox{ (for fermions)}.
\end{align*}

To tackle the problem~(\ref{eq2DShroedingerEquationBasic})--(\ref{eqBoundaryConditionAtInfinity}), along
with interaction potential~(\ref{eqPotentialU}) we apply the numerical scheme, that we applied to the quantum anisotropic scattering in a plane~\cite{Koval_2014}. 

The following wave function expansion is applied:
\begin{equation}
\label{eqJETPFullPsiExpansion}
\Psi \left( {\rho ,\phi } \right)
\approx
\frac{1}{\sqrt \rho 
} {\sum\limits_{m=-M}^M \sum\limits_{j=0}^{2M}{\xi _m (\phi )\xi 
_{mj}^{-1} \psi _j (\rho )} },
\end{equation}
where $\xi _{mj}^{-1} =\frac{2\pi }{2M +1}\xi _{jm}^\ast =\frac{\sqrt{2\pi}
 }{2M +1}e^{-im(\phi _j - \pi) }$~--- is the inverse matrix to the square matrix $\left({2M +1} \right)\times \left( {2M +1} \right)$ ${\xi _{jm} =\xi _m (\phi _j )}$, that is defined on the uniform angular grid ${\phi_j =\frac{2\pi j}{2M+1}}(\mbox{where }j=0,1,...,2M)$. In the angular grid's nodes ${\phi _j}$: $\Psi \left( {\rho ,\phi_j} \right) \approx \psi_j(\rho)/\sqrt{\rho}$.
$\xi_m (\phi )=\frac{(-1)^m}{\sqrt {2\pi } }e^{im\phi }$ are the eigenfunctions of the operator $h^{(0)}(\phi )=\tfrac{\partial^2}{\partial \phi^2}$ and serve as a basis of functions for the wave function expansion over the angular variable. 
 
In representation~(\ref{eqJETPFullPsiExpansion}), the 2D Schr\"{o}dinger equation transforms in the system of $(2M+1)$ coupled second-order differential equations:
\begin{align}
 \label{eq10}
 \frac{d^2\psi _j (\rho )}{d\rho ^2}+\frac{2\mu }{\hbar ^2}\left( {E-U(\rho
 ,\phi _j )
 +\frac{\hbar ^2}{8\mu \rho ^2}} \right)\psi _j (\rho
 )+\nonumber \\
-\frac{1}{\rho ^2}\sum\limits_{j'} \sum\limits_{j''=-M}^M j''^2 \xi _{jj''} \xi _{j''j'}^{-1} 
\psi _{j'}(\rho) =0.
\end{align}
The seven-point finite difference approximation of six-order accuracy is used for the derivatives discretization. An obtained on each iteration matrix problem is tackled with the matrix modification of the sweep algorithm for the band matrix.

The original numerical algorithm~\cite{Koval_2014} was modified. For a differential grid over a radial variable ($\{\rho_k\}; \mbox{ } ({k=0,1,...,N})$) we use the non-uniform grid: 
\begin{align}
    \rho_{k} = & \rho_{0} + (\rho_{N}-\rho_{0}) t_{k}^2,  \\
    t_{k} = & k/N; t_{k} \in [0,1].
\end{align}
and which is similar to the quasi-uniform grids~\cite{Kalitkin_2005_english}.
This modification  allows us to substantially decrease the number of differential grid nodes $N$ over the radial variable $\rho$ and to increase the algorithm's convergence rate, which has saved a lot of computation time.

\section{Results}
\label{sec:Results}

\begin{figure}[tbp]
 \centerline{\includegraphics[width=0.7\linewidth]{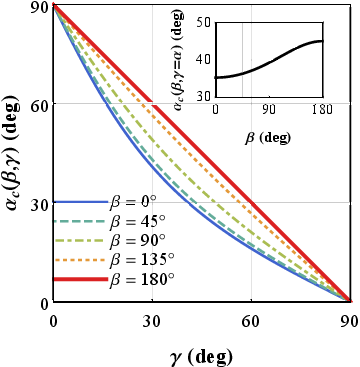}}
 \caption{The dependence of the critical tilt angle $\alpha_{c}(\beta,\gamma)$ of the dipole $\bf d_1$ on the rotation angle $\beta$ and on the tilt angle $\gamma$ of the dipole $\bf d_2$ (see the scheme in Fig.~\ref{figSchemeOfTwoDipoles}). The inset presents the critical tilt angle $\alpha_{c}(\beta,\gamma)$ as a function of the rotation angle $\beta$ for the case of the dipoles with equal tilt angles ($\gamma=\alpha$).} 
 \label{figCriticalAngle}
 \end{figure}

\begin{figure*}[tb]\begin{minipage}[t]{1.0\linewidth}
        \begin{minipage}[t][][t]{0.47\linewidth}
            \begin{overpic}[width=\linewidth]{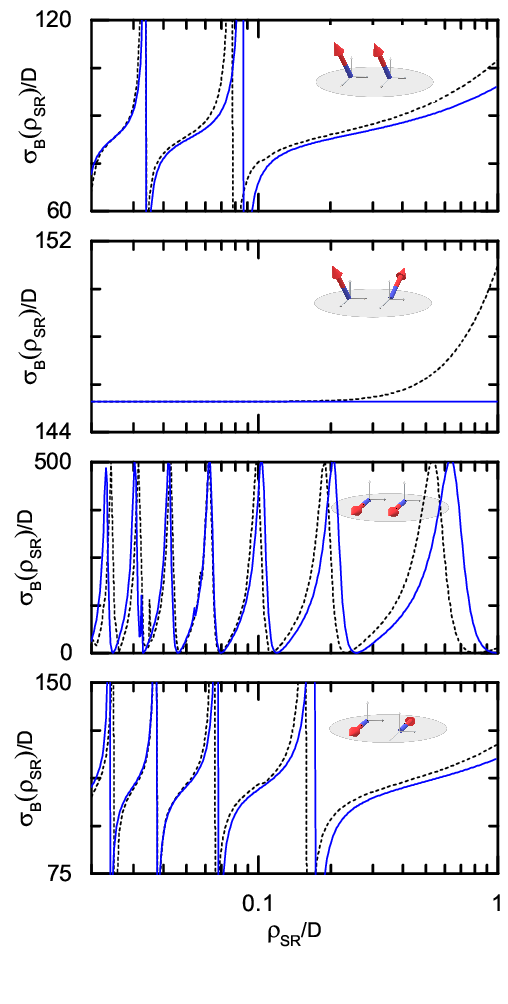}
                \put(45,94){
                \Large{\textbf{(\textit{a})}}
                }
                \put(45,71){
                \Large{\textbf{(\textit{b})}}
                }
                \put(46,49){
                \Large{\textbf{(\textit{c})}}
                }
                \put(45,26){
                \Large{\textbf{(\textit{d})}}
                }
            \end{overpic}
            \caption{
                The dependence of the total cross section of the dipolar low-energy scattering $\sigma_{B}$  of \textit{identical bosons} on the SRI radius~$\rho_{SR}$ at collisions of two aligned $\beta=0^{\circ}$ ($a,c$) and unaligned $\beta=180^{\circ}$ ($b,d$) dipoles (at $\gamma=\alpha$), tilted to the angle $\alpha=45^{\circ}$ ($a,b$), as well as for a limiting case of dipoles lying in the scattering plane $\alpha=90^{\circ}$ ($c,d$). The results obtained with the use of the hard wall potential are marked with a black dashed line, whereas those obtained by the use of the Lennard-Jones potential~--- by the blue solid line.
            }
            \label{figFullCrossSectionBosons}
        \end{minipage}
        \hfill
        \begin{minipage}[t][][t]{0.47\linewidth}
            \begin{overpic}[width=\linewidth]{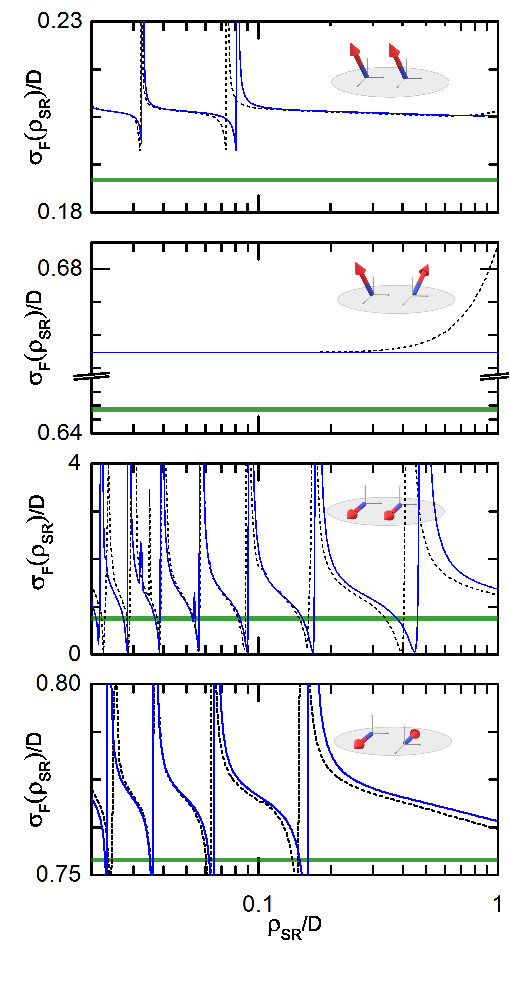}
                \put(45,94){
                \Large{\textbf{(\textit{a})}}
                }
                \put(45,71){
                \Large{\textbf{(\textit{b})}}
                }
                \put(45,49){
                \Large{\textbf{(\textit{c})}}
                }
                \put(45,26){
                \Large{\textbf{(\textit{d})}}
                }
            \end{overpic}
            \vfill
            \caption{
                The same as in Fig.~\ref{figFullCrossSectionBosons} for \textit{identical fermions}.
The Born approximation is marked by a green bold line. 
            }
             \label{figFullCrossSectionFermions} 
        \end{minipage}
    \end{minipage}
\end{figure*}

\subsection{Critical (magic) angle for scattering of arbitrarily oriented dipoles}
\label{sec:Results:criticalAngle}

Here we consider the reshaping of the dipole-dipole interaction potential~$V_{dd}$ with increasing the tilt angle $\alpha$ of the dipole $\bf d_1$, with respect to the direction of the dipole $\bf d_2$: with the tilt angle $\gamma$ and rotation angle $\beta$ between the planes $Zd_1$ and $Zd_2$ (see Fig.~\ref{figSchemeOfTwoDipoles}).

Domains of the attractive dipolar interaction arise around the points $\phi'$, that are defined by the expression: $\left.\tfrac{\partial V_{dd}(\rho,\phi)}{\partial \phi} \right|_{\phi'}=0$.
The \textit{critical tilt angle} $\alpha_c(\beta,\gamma)$ is defined as the angle $\alpha$ (see Fig.~\ref{figSchemeOfTwoDipoles}), above which values of $V_{dd}$ potential become negative in the points $\phi = \phi'$: $V_{dd}(\rho,\phi)<0$. Thus, the condition:
\begin{equation}
V_{dd}(\rho,\phi')=0,
\end{equation}
determines the dependence of the critical tilt angle $\alpha=\alpha_{c}(\beta,\gamma)$ of the dipole $\bf d_1$ on the rotation angle $\beta$ and the tilt angle $\gamma$ of the dipole $\bf d_2$:
\begin{equation}
\label{eqCriticalAngleForArbitraryBetaGamma}
\alpha_{c}(\beta,\gamma) = \arctan \left( \frac{2 \cot(\gamma)}{3+\cos(\beta)} \right), 
\end{equation}
presented in Fig.~\ref{figCriticalAngle}. The critical tilt angle $\alpha_{c}(\beta,\gamma)$ increases as $\beta \to 180^{\circ}$ (e.g. at $\gamma=45^{\circ}$ the angle $\alpha_{c}$ increases from $26.56^{\circ}$ to $45^{\circ}$). It should be noted, that at $\beta=180^{\circ}$ the critical tilt angle can be found from a plain ratio $\alpha_{c}(\beta,\gamma) = 90^{\circ} - \gamma$, which is indicated in Fig.~\ref{figCriticalAngle} by the solid red line.

When the dipoles' tilt angles are equal ($\gamma=\alpha$) and the second dipole is rotated along the $Z$ axis, the critical tilt angle as a function of the rotation angle $\beta$ has the form:
\begin{equation}
\label{eqCriticalAngleForArbitraryBeta}
\alpha_c(\beta,\gamma)=\arctan\sqrt[]{\frac{2}{3+\cos(\beta)}},
\end{equation}
(see the inset of Fig.~\ref{figCriticalAngle}).
When one considers the aligned dipoles case $\gamma=\alpha$ and $\beta=0^{\circ}$, the expression~(\ref{eqCriticalAngleForArbitraryBeta}) reproduces the known value of the critical (magic) angle ${\alpha_c(\beta,\gamma) = \arctan{\frac{1}{\sqrt[]{2}}} \approx 35.3^{\circ}}$ or ${90^{\circ} - \alpha_c(\beta,\gamma) = 54.7^{\circ}}$ (if the tilt angle is defined with respect to the plane $XY$), as previously mentioned~\cite{Pfau_2002,Ticknor_2011,Macia_2011}.
The maximal value of the critical angle $\alpha_c(\beta,\gamma)$ at $\gamma=\alpha$ is equal to $45^{\circ}$ and it is reached at $\beta=180^{\circ}$.

\begin{figure*}[htb]\centering
\vfill
\begin{minipage}[h]{0.49\linewidth}
\begin{overpic}[height=6.7cm]{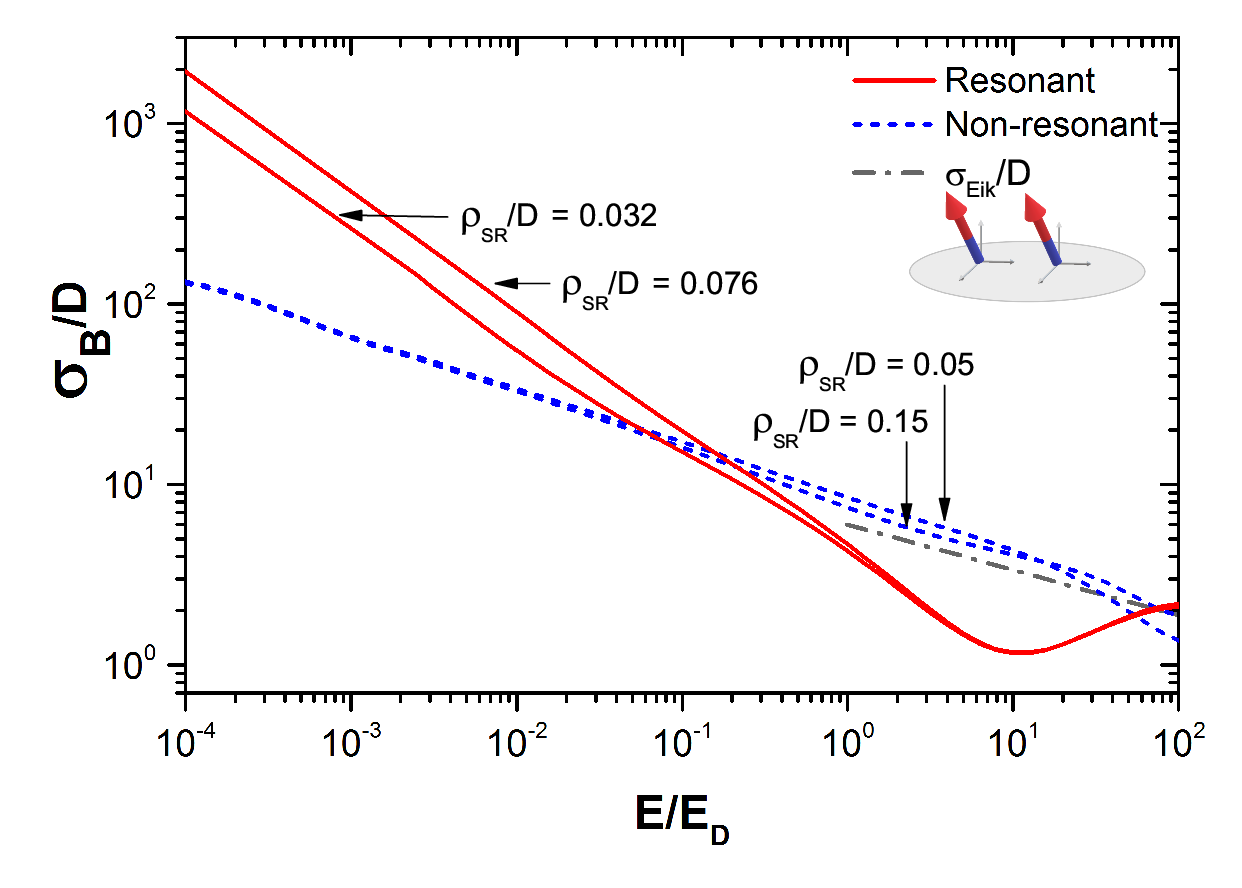}
\put(84,43)
{
\textbf{(\textit{a})}
}
\end{overpic}
\end{minipage}
\hfill
\begin{minipage}[h]{0.49\linewidth}
\begin{overpic}[height=6.7cm]{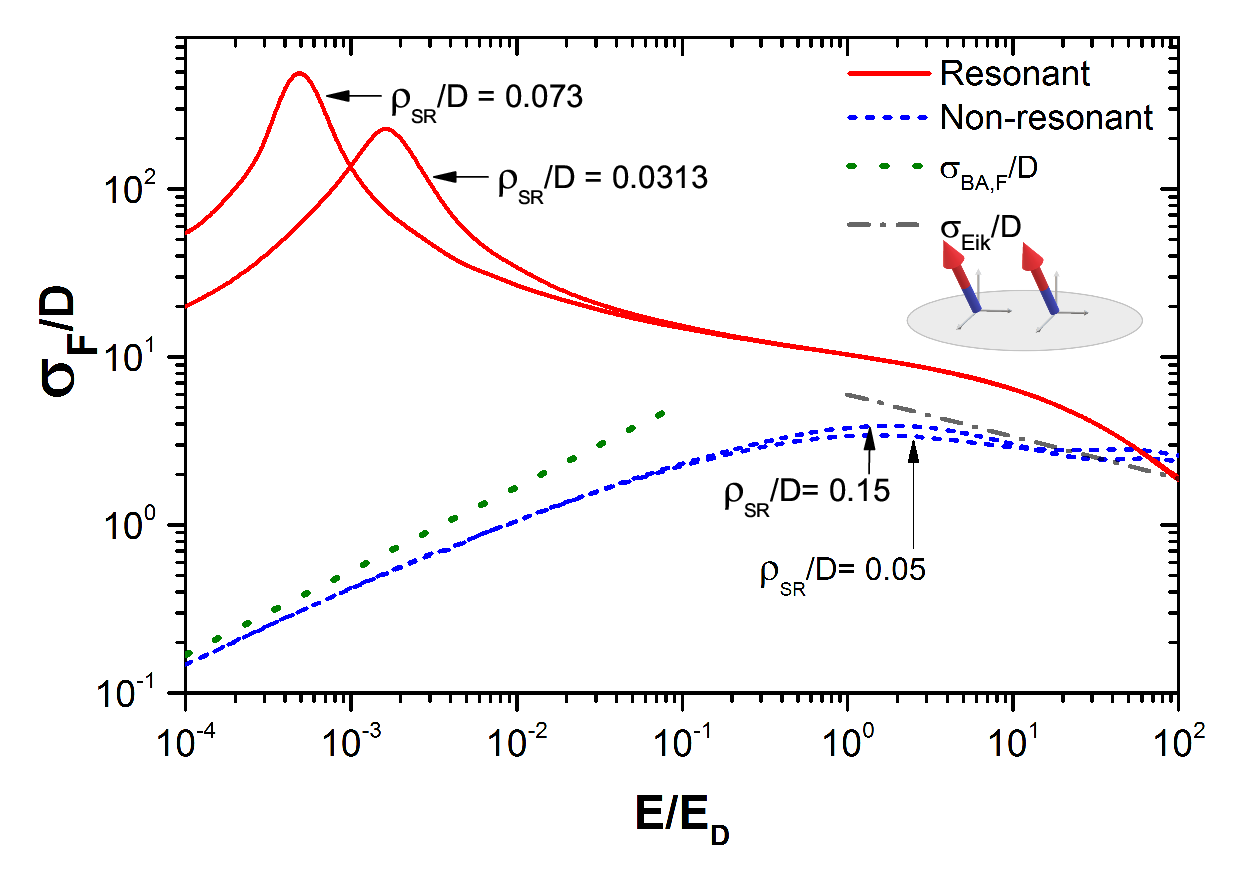}
\put(84,43)
{
\textbf{(\textit{c})}
}
\end{overpic}
\end{minipage}
\vfill
\begin{minipage}[h]{0.49\linewidth}
\begin{overpic}[height=6.7cm]{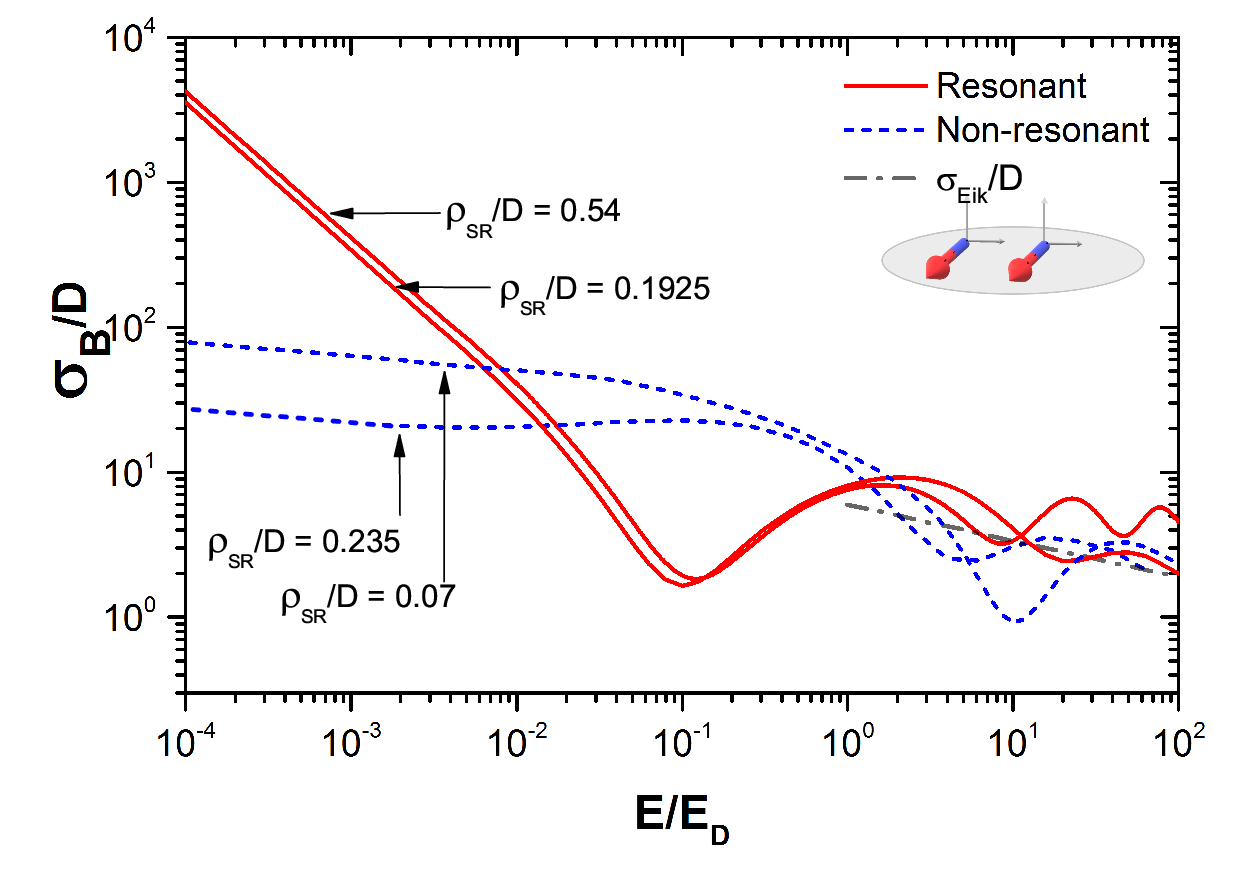}
\put(84,43)
{
\textbf{(\textit{b})}
}
\end{overpic}
\end{minipage}
\hfill
\begin{minipage}[h]{0.49\linewidth}
\begin{overpic}[height=6.7cm]{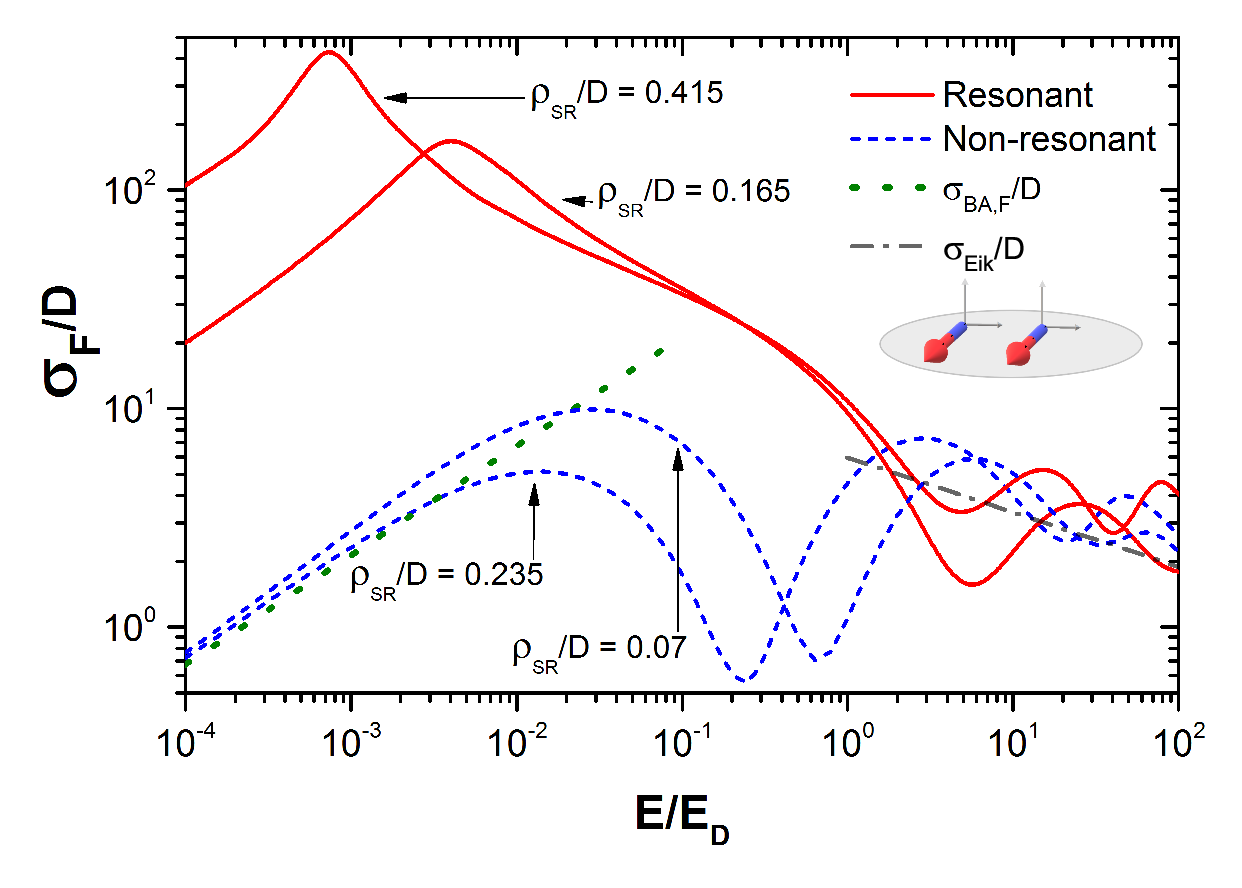}
\put(84,43)
{
\textbf{(\textit{d})}
}
\end{overpic}
\end{minipage}
\vfill
\caption{The energy dependencies of the total cross section of the dipolar scattering of identical bosons $\sigma_{B}(E)$ ({$\alpha = 45^{\circ}$(a), $90^{\circ}$(b)}) and identical fermions $\sigma_{F}(E)$ ($\alpha = 45^{\circ}$(c), $90^{\circ}$(d)) for aligned dipoles configuration $\beta=0^{\circ}$ and $\gamma=\alpha$. The tilt angle exceeds the critical angle ${\alpha >\alpha_c}$ ($\alpha_c = 35.3^{\circ}$) for such dipole mutual orientations. The curves corresponding to the resonance points in Figs.~\ref{figFullCrossSectionBosons} and \ref{figFullCrossSectionFermions} are indicated by a red solid line, the non-resonant curves by a blue dashed line; the Born approximation by a green dotted line, the eikonal approximation by a gray dashed line. 
} 
 \label{figOverEnergy}
 \end{figure*}

\subsection{SRI-induced resonances in low-energy 2D scattering of boson and fermion dipoles}
\label{sec:Results:SRIinducedResonances}

We numerically calculated the accurate values of the total cross section $\sigma(\rho_{SR})$ of a 2D scattering  as a function of a SRI radius $\rho_{SR}$ at the low collision energy ${E = 5 \times 10^{-4}E_D}$.  
The obtained results for the identical boson and identical fermion scattering are illustrated in Figs.~\ref{figFullCrossSectionBosons} and~\ref{figFullCrossSectionFermions}, respectively.

Fig.~\ref{figFullCrossSectionBosons} presents the dependence of the total cross section of the dipolar scattering of \textit{identical bosons} on the SRI radius~$\rho_{SR}$ in the collision of two \textit{aligned} $\beta=0^{\circ}$ ($a,c$) and \textit{unaligned} $\beta=180^{\circ}$ ($b,d$) dipoles (at $\gamma=\alpha$), tilted at the angle $\alpha=45^{\circ}$ ($a,b$), as well as for a limiting case of dipoles lying in the scattering plane $\alpha=90^{\circ}$ ($c,d$). The results for the hard wall potential are marked by a black dashed line, whereas for the Lennard-Jones potential by the blue solid line.
The distinct narrow resonances in the dependence of the calculated total cross section $\sigma_B(\rho_{SR})$ for bosonic dipoles occur for the scattering of the two \textit{aligned} dipoles ($\beta=0^{\circ}$; $\gamma = \alpha$) at tilt angles $\alpha = 45^{\circ}$, presented in Fig.~\ref{figFullCrossSectionBosons}(a). The number of resonances in the cross section dependence on the $\rho_{SR}$ is quadrupled with increasing tilt angle $\alpha$ from $45^{\circ}$ up to $90^{\circ}$. The calculated dependence of the $\sigma_B(\rho_{SR})$ for the two dipoles, that are directed in the scattering plane $XY$ ($\alpha=90^{\circ}$) is presented in Fig.~\ref{figFullCrossSectionBosons}(c).
The analysis of Fig.~\ref{figFullCrossSectionBosons}(c) shows, that as $\rho_{SR}$ decreases the resonances increases in number while simultaneously becoming narrower.  

The cross section does not depend on the SRI potential and there are no dipolar scattering resonances at $\alpha \leq \alpha_{c}(\beta,\gamma)$ and $\rho_{SR}/D \ll 1$, that is shown in Fig.~\ref{figFullCrossSectionBosons}(b) for $\alpha = 45^{\circ}$ ($\gamma = \alpha$). Thus, the rotation of the dipole moment vector ${\bf d}_2$ around the $Z$ axis by $\beta \to 180^{\circ}$ causes the narrowing of the domains, where the dipolar potential is attractive, and the number of the resonances decreases until they disappear at $\beta \to 180^{\circ}$.
It should be noted, that the resonances emerge also in the case of oppositely oriented dipoles ($\beta = 180^{\circ}$) lying in the scattering plane at $\alpha \to 90^{\circ}$ ($\gamma = \alpha$,$\alpha > \alpha_c$),  as illustrated in Fig.~\ref{figFullCrossSectionBosons}(d).
The increase of tilt angles $\alpha,\gamma$ of dipoles and arising of domains with attractive dipolar potential leads to the strong dependence of the dipolar scattering cross section on the radius $\rho_{SR}$ of the SRI potential.

The distinguishable particles scattering in low-energy limit has also been investigated and the obtained results are almost matching within a factor of 4 with the ones for the scattering of identical bosons, due to the $s$-wave contribution, that grows with an energy decrease.

The analysis of our calculations results has showed, that the fermion scattering is significantly different from the boson scattering at low energies. As expected according to the Wigner threshold law~\cite{Sadeghpour_2000} the fermion scattering is suppressed compared with the boson scattering. The calculated total cross section of dipolar scattering of \textit{identical fermions} as a function of~$\rho_{SR}$ is presented in Fig.~\ref{figFullCrossSectionFermions} for equal angles $\gamma=\alpha$ of dipoles, tilted to the angle $45^{\circ}$ for aligned $\beta=0^{\circ}$ ($a$) and unaligned $\beta=180^{\circ}$ ($b$) configurations, and also for the limiting case of two dipoles lying in the scattering plane $\alpha=90^{\circ}$ for parallel $\beta=0^{\circ}$ ($c$) and antiparallel $\beta=180^{\circ}$ ($d$) dipole moments.  
The comparative analysis of Fig.~\ref{figFullCrossSectionBosons} and Fig.~\ref{figFullCrossSectionFermions} demonstrates, that for the case of fermion collisions, the amplitudes of the resonances are two orders of magnitude smaller, than those for bosons, in the case of low energy of the collision. 

The shapes of resonant curves in Figs.~\ref{figFullCrossSectionBosons}(c)~and~\ref{figFullCrossSectionFermions}(c) for the particular case of 2D scattering of aligned dipoles lying in the scattering plane ($\alpha=90^{\circ}$) have the similar profile with the results of Ref.~\cite{Cavagnero_2009}, obtained for the low-energy 3D dipolar scattering (see Fig.3 of Ref.~\cite{Cavagnero_2009}). In our opinion, this fact of the similar profile is due to the high probability of the ``head-to-tail'' dipole collisions.

We have also calculated the dependencies $\sigma_{B}(\rho_{SR}),\sigma_{F}(\rho_{SR})$ when using the realistic Lennard-Jones potential as $V_{SR}$. The results are presented in Figs.~\ref{figFullCrossSectionBosons} and \ref{figFullCrossSectionFermions} and are marked by a blue solid line.  
The Lennard-Jones potential models short-range repulsion more physically, in the sense of introducing correlations between the different partial waves short-range phases, which shift narrow resonances in high partial waves, as seen in Figs.~\ref{figFullCrossSectionBosons} and~\ref{figFullCrossSectionFermions}. But the resonances' structure remains qualitatively the same as when using the hard wall potential, because the resonances are due to the $s$-wave ($p$-wave) dominance in the scattering of bosons (fermions). The comparison of the calculation results, which were obtained by using the hard wall potential and the Lennard-Jones potential, shows their equal efficiency. The results of calculations are qualitatively and quantitatively similar. So we make a conclusion about a good applicability of the hard wall potential for the 2D dipolar scattering characteristic calculations and hereafter we present the results, which were obtained using the hard wall potential.

We reveal, that the 2D dipolar scattering cross section dependencies on the SRI radius for both bosons and fermions are strongly changed for different mutual orientations of arbitrarily directed dipoles. Figs.~\ref{figFullCrossSectionBosons} and~\ref{figFullCrossSectionFermions} illustrate this fact.

\subsection{Energy dependencies of cross section of bosonic and fermionic dipoles scattering in plane}
\label{sec:Results:EnergeticDependencies}

The energy dependencies of the cross section of the dipolar scattering in a plane were studied in Ref.~\cite{Ticknor_2009} for the isotropic repulsive case of dipolar interaction. Here we present the results of the calculations of the energy dependencies for the anisotropic dipolar interaction potential $V_{dd}$.
The energy dependencies of the total cross section of the dipolar scattering of identical bosons $\sigma_{B}(E)$ ({$\alpha = 45^{\circ}$(a), $90^{\circ}$(b)}) and identical fermions $\sigma_{F}(E)$ ($\alpha = 45^{\circ}$(c), $90^{\circ}$(d)) for aligned dipoles configuration $\beta=0^{\circ}; \gamma=\alpha$ are illustrated in Fig.~\ref{figOverEnergy}. The tilt angle exceeds the critical angle ${\alpha >\alpha_c}$ ($\alpha_c = 35.3^{\circ}$) for such dipole mutual orientations. The curves corresponding to the resonance points in Figs.~\ref{figFullCrossSectionBosons} and \ref{figFullCrossSectionFermions} are indicated by a red solid line; the non-resonant curves by a blue dashed line; the Born approximation by a green dotted line, the eikonal approximation by a gray dashed line.

The analysis of the dependencies shows that in a low-energy limit, the cross section of resonant cases is at least an order of magnitude greater than the values for non-resonant cases. 
The cross section of identical bosons scattering increases in the  $E \to 0$ limit in both cases due to the $s$-wave, which is caused by the divergence in 2D space~\cite{Simon_1976}. It should be noted, that in the vicinity of resonances the cross section is an order of magnitude greater than those at the absence of resonances.
All resonant curves of the cross section $\sigma_{F}(E)$ for the dipolar scattering of fermions demonstrate a peak shape in the low-energy limit, in contrast to the non-resonant curves, that monotonically decrease. These peaks shift to the lower energies with the growing value of its maximums at an increase of $\rho_{SR}$. The resonances for fermions are narrower than for bosons, due to potential barriers for high partial waves, that suppress the partial cross section in the low-energy limit.

The boson (fermion) dipoles 2D scattering cross section in the absence of resonances increases (decreases) in the low-energy limit in contrast to the 3D scattering, where the cross section in the absence of resonances has the form of a plateau in the low-energy limit for both bosons and fermions (see Fig.~2(a,b) in Ref.~\cite{Cavagnero_2009} or Fig.~1 in Ref.~\cite{bohn2009quasi}).

We also showed that the mutual orientation of dipoles strongly impacts the form of the energy dependencies. Thus, at an increase of the angle $\alpha$ from $45^{\circ}$ to $90^{\circ}$ (see Fig.~\ref{figOverEnergy}) the resonant and non-resonant dependencies $\sigma_{B}(E), \sigma_{F}(E)$ begin to oscillate, unlike the 3D dipolar scattering~\cite{Cavagnero_2009,bohn2009quasi}, where there is no such oscillations.

\subsection{Born approximation}
\label{sec:Results:BornApproximation}
 
For weak dipole moments, the Born approximation (BA)~\cite{Lapidus_1982} was used to estimate the ultracold dipoles scattering amplitude in Refs.~\cite{oldziejewski2016properties, kanjilal2008low, ronen2006dipolar, Ticknor_2009, Ticknor_2011,bohn2014differential}. A good agreement with close-coupling calculations~\cite{oldziejewski2016properties, kanjilal2008low, ronen2006dipolar, Ticknor_2009, Ticknor_2011,bohn2014differential} was obtained away from resonances. However, in the vicinity of the resonances the Born approximation is not valid and one has to numerically calculate the scattering cross sections~\cite{BOOK_Sakurai_ModernQuantumMechanics_2017}.

\begin{figure}[t]\centering
\begin{overpic}[width=8.6cm]{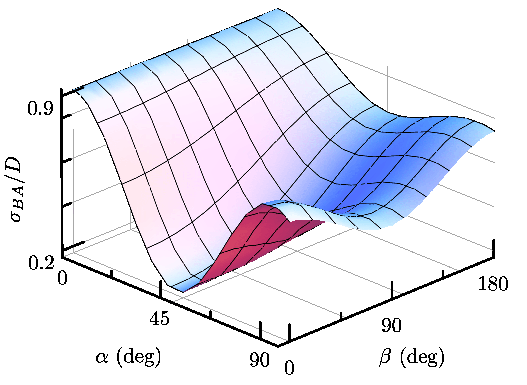}
\end{overpic}
\caption{The dependence of the total cross section for dipolar scattering of fermions on the tilt angle $\alpha$ in degrees (at $\gamma=\alpha$) and on the rotation angle $\beta$ in degrees obtained within the Born Approximation.} 
\label{figBornApproximationForFermions}
\end{figure}

We generalize the expression of the total cross section of aligned dipoles, obtained within the Born approximation in Ref.~\cite{Ticknor_2011} for the long-range part of the potential~(\ref{eqPotentialVDipoleDipoleInPolarCoordinates}) by a summation of three series of partial cross section, for the case of \textit{arbitrarily oriented} dipoles (\textit{identical fermions}):
\begin{multline}
\label{eqBornApproximationFull}
\sigma_{BA}/D = 8 \sqrt{2 E/E_D}
\Biggl[ 
    \pi (\pi - 2)  
    \Bigl( 
        \sin(\alpha) \sin(\gamma) \cos(\beta) + \Bigr. \\ \Bigl. + \cos(\alpha) \cos(\gamma) - \frac{3}{2} \sin(\alpha) \sin(\gamma) \cos(\beta)
    \Bigr)^2  \\ + \left(  
    \pi (\pi + 2) + \frac{32}{9} -\frac{63520}{3969} \right)
    \left( 
        \frac{3}{4} \sin(\alpha) \sin(\gamma)
    \right)^2
\Biggr].
\end{multline}
The dependence~(\ref{eqBornApproximationFull}) of the BA on the the tilt angle $\alpha$ (at $\gamma=\alpha$) and the rotation angle $\beta$ at the energy $E = 5 \times 10^{-4}E_D$ is illustrated in Fig.~\ref{figBornApproximationForFermions}. 

As seen in Fig.~\ref{figFullCrossSectionFermions}, the $\sigma_{F}$ obtained within the Born approximation (BA), that is marked with a bold green line, roughly agrees with the calculated values of $\sigma_{F}({\rho_{SR}})$ away from resonances, but it is inapplicable when the resonant structure of the $\sigma_{F}({\rho_{SR}})$ dependence is dense. The BA predicts the same cross section values for the $\beta=0^{\circ}$ and $\beta=180^{\circ}$ at $\alpha=90^{\circ}$. On the basis of the analysis of Fig.~\ref{figBornApproximationForFermions}, one could expect that the scattering cross section should be less dependent on the tilt angle $\alpha$ as $\beta \to 180^{\circ}$ and should approach the value of $\alpha=0^{\circ}$ case, which is confirmed only for the $\alpha<\alpha_c$. 
Thus, the calculated dependencies show, that the BA~(\ref{eqBornApproximationFull}) is applicable to the two-dimensional dipolar scattering of fermions when the tilt angle of the dipoles is less than critical angle $\alpha<\alpha_c$, when the DDI potential does not support bound states.

The BA is not directly applicable to the scattering of identical bosons or distinguishable particles, due to a divergence of the $s-$wave contribution in the region of the short-range potential, which must be explicitly accounted, since there is no centrifugal barrier for the $s-$wave. The divergence could be avoided by the use of the pseudopotential with a ``potential strength''~\cite{ronen2006dipolar}. Although, in order to obtain the ``potential strength'' it is necessary to numerically solve the full two-dimensional Shr\"{o}dinger equation, so it is useful only for the following many-body calculations. That is why we compare the BA only with the scattering of fermions. 

In the low-energy limit the values of $\sigma_{BA}(E)$, obtained within the BA, marked by a green dotted line in Fig.~\ref{figOverEnergy}($c$,$d$), agrees well with the non-resonant dependencies $\sigma_F(E)$ for the fermion scattering. Note, that there is no such agreement for the resonant curves in Fig.~\ref{figOverEnergy}.

\subsection{Eikonal approximation}
\label{sec:Results:EnergeticDependencies:EikonalApproximation}

In the high energy regime ${E \gg E_D}$, the semi-classical eikonal approximation~\cite{Ticknor_2009,Ticknor_2011,adhikari2008semiclassical}:
\begin{equation}
\label{eqEikonalApproximation}
\sigma_{Eik}/D = \frac{4\sqrt{\pi}}{(2 E/E_D)^{1/4}},
\end{equation}
is applicable to estimate the scattering cross section. For $E > 50 E_D$ the dependencies of the 2D dipolar scattering cross section on the collision energy of bosons, $\sigma_{B}(E)$, and fermions, $\sigma_{F}(E)$, oscillate around the curve of the eikonal approximation $\sigma_{Eik}(E)$, marked by a grey dot-dashed line in Fig.~\ref{figOverEnergy}($c$,$d$).  
The dipolar scattering cross sections, obtained within the eikonal approximation, are in an excellent agreement with presented in Fig.~\ref{figOverEnergy} numerically calculated cross sections of dipolar scattering $\sigma_{B}$ and $\sigma_{F}$, that correspond to the absence of resonances, at $E > 50 E_D$. 
This is explained as follows: the more partial waves are involved in the scattering with an increase of the collision energy, the more the scattering cross section approaches to semiclassical estimates. 
However, the cross sections of dipolar scattering of bosons and fermions obtained numerically, which correspond to the occurrence of resonances, are strongly different from those of $\sigma_{Eik}$, which indicates a significant impact of the $s-$ and $p-$waves on the scattering of dipoles even for the high energies $E \gg E_D$.

\subsection{Impact of SRI on high-energy 2D scattering of arbitrarily oriented dipoles}

\begin{figure*}[htbp]\centering
\vfill
\hfill
\begin{minipage}[h]{0.49\linewidth}
\begin{overpic}[height=7cm]{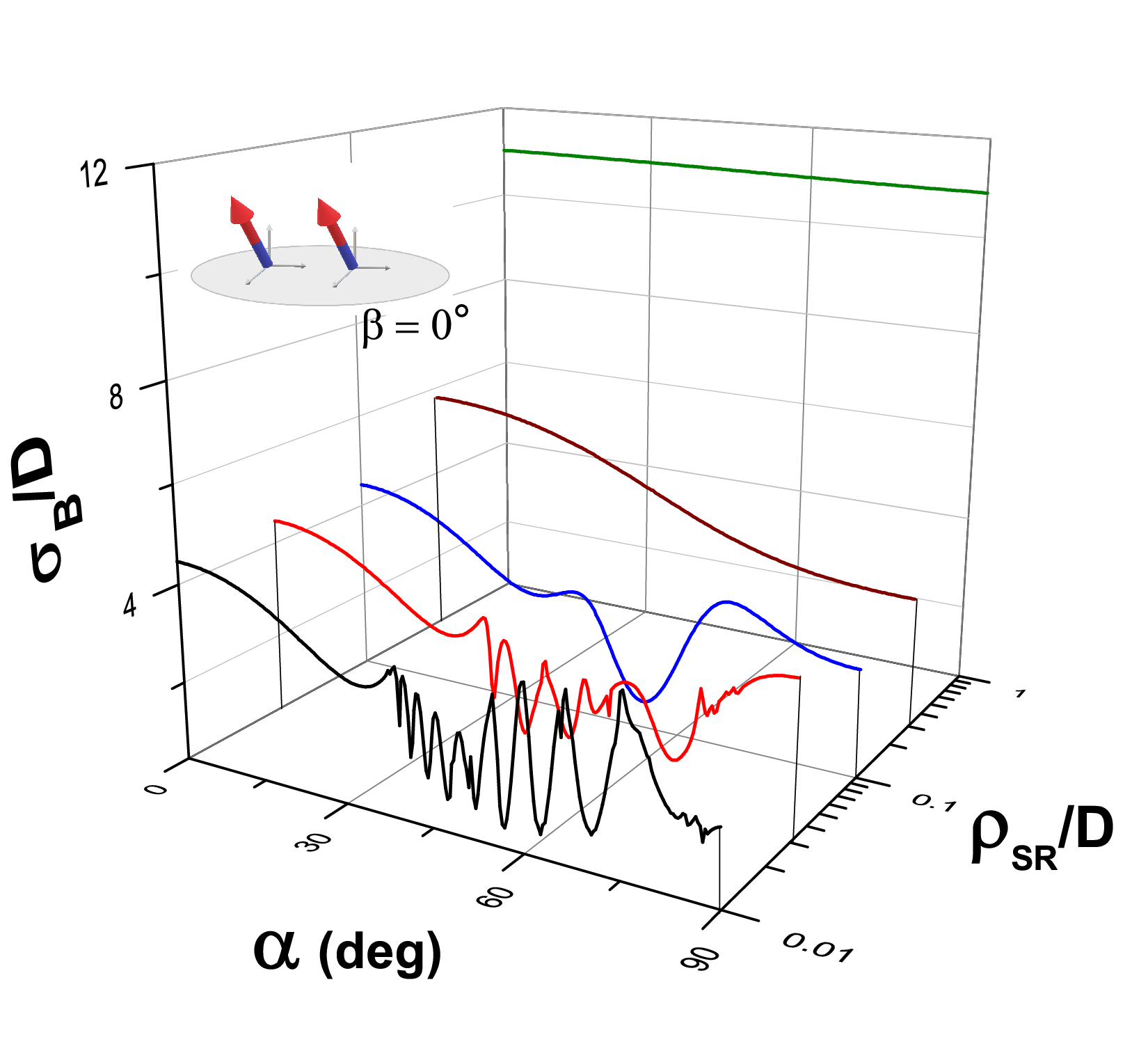}
\put(76,67){\Large{\textbf{(\textit{a})}}}
\end{overpic}
\end{minipage}
\hfill
\begin{minipage}[h]{0.49\linewidth}
\begin{overpic}[height=7cm]{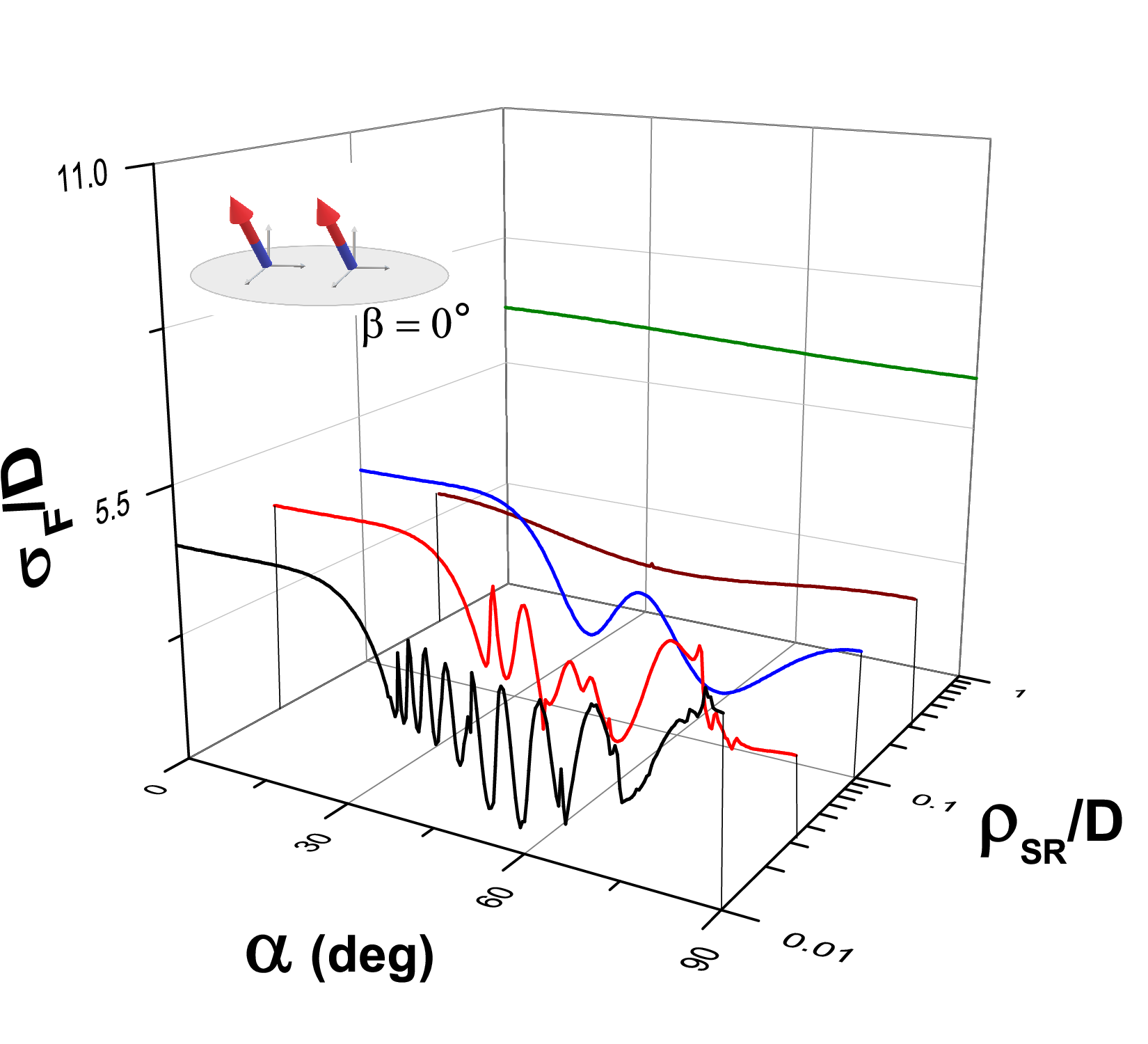}
\put(76,67){\Large{\textbf{(\textit{b})}}}
\end{overpic}
\end{minipage}
\hfill
\vfill
\hfill
\begin{minipage}[h]{0.49\linewidth}
\begin{overpic}[height=7cm]{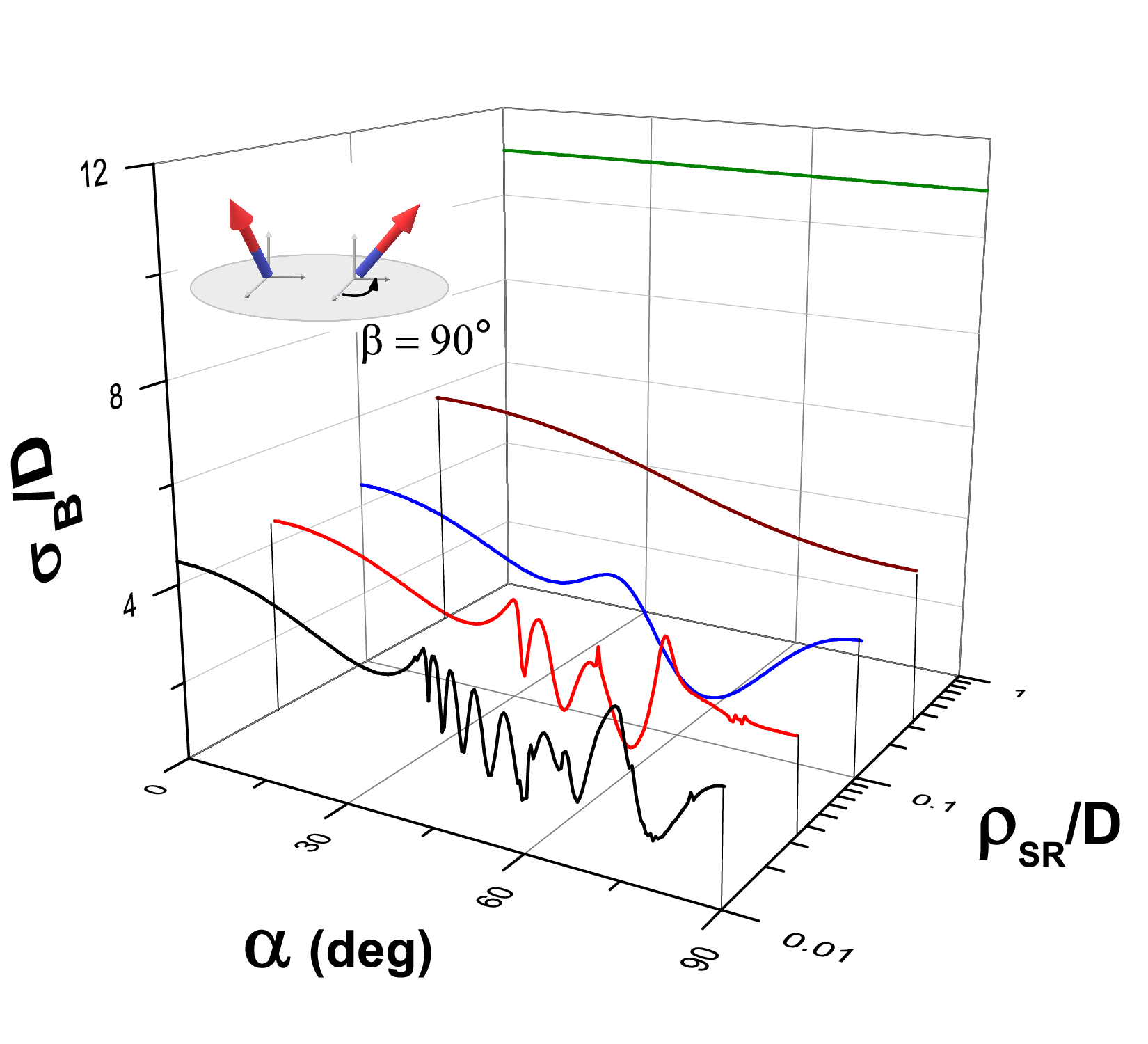}
\put(76,67){\Large{\textbf{(\textit{c})}}}
\end{overpic}
\end{minipage}
\hfill
\begin{minipage}[h]{0.49\linewidth}
\begin{overpic}[height=7cm]{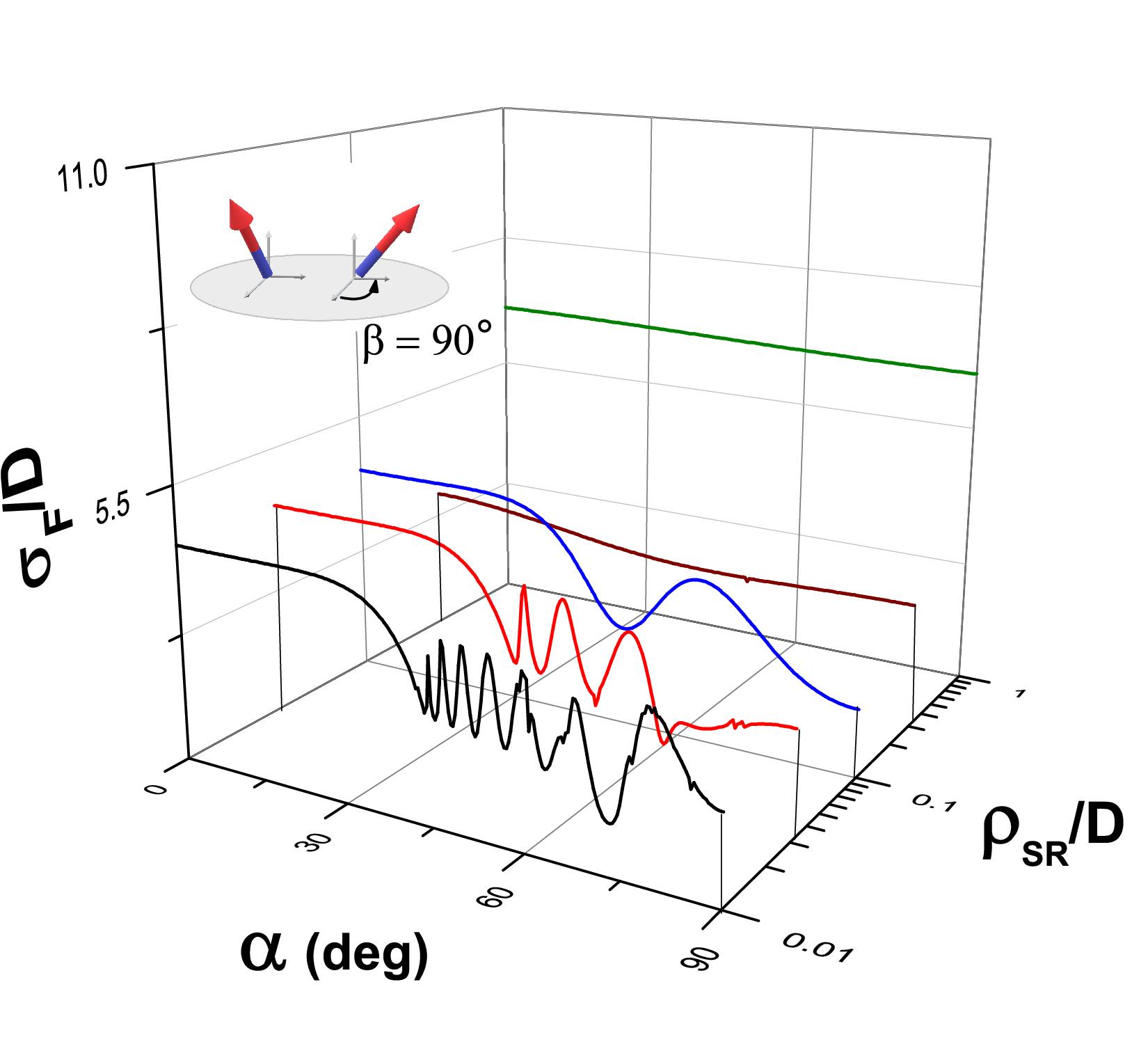}
\put(76,67){\Large{\textbf{(\textit{d})}}}
\end{overpic}
\end{minipage}
\hfill
\vfill
\hfill
\begin{minipage}[h]{0.49\linewidth}
\begin{overpic}[height=7cm]{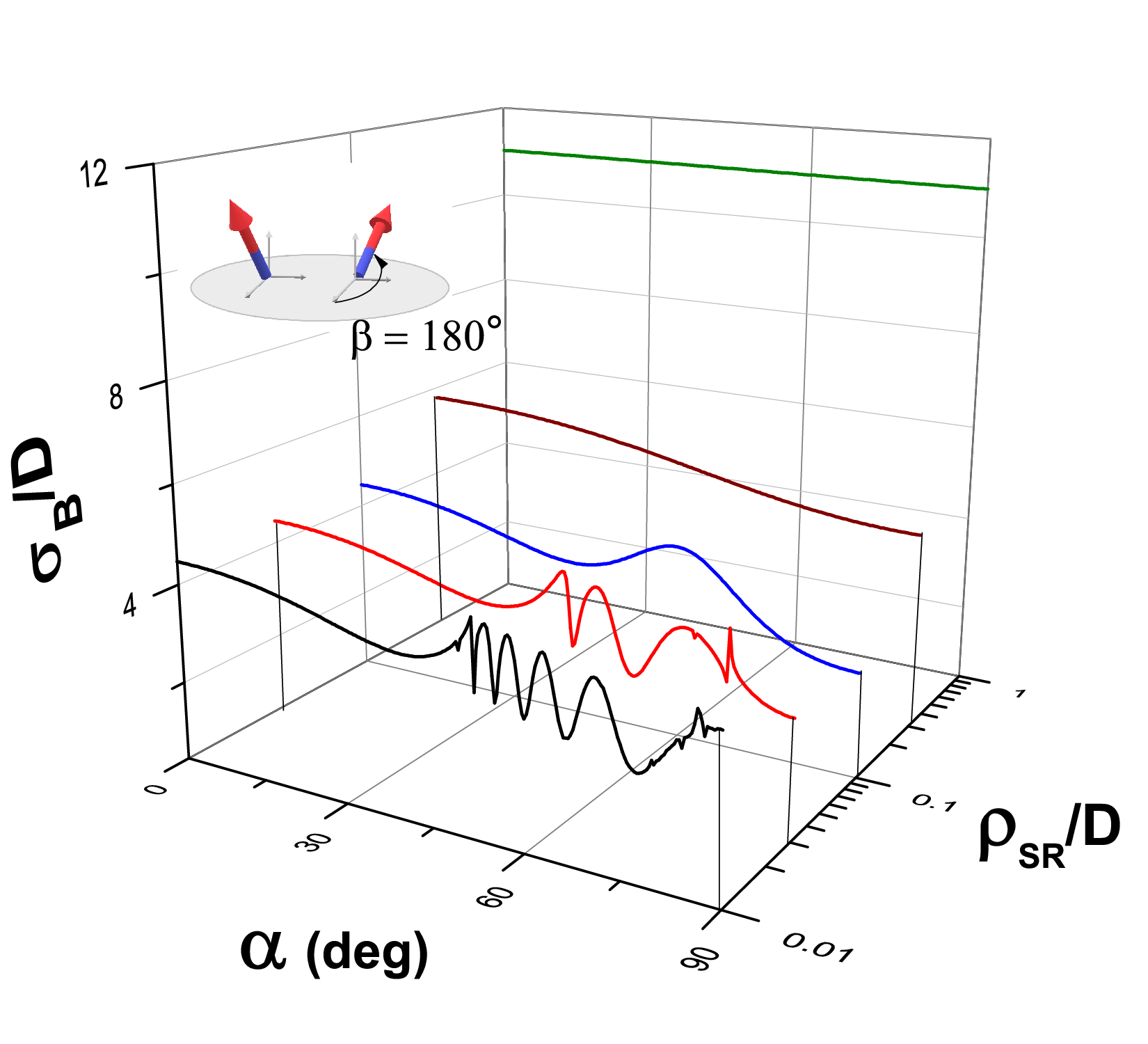}
\put(76,67){\Large{\textbf{(\textit{e})}}}
\end{overpic}
\end{minipage}
\hfill
\begin{minipage}[h]{0.49\linewidth}
\begin{overpic}[height=7cm]{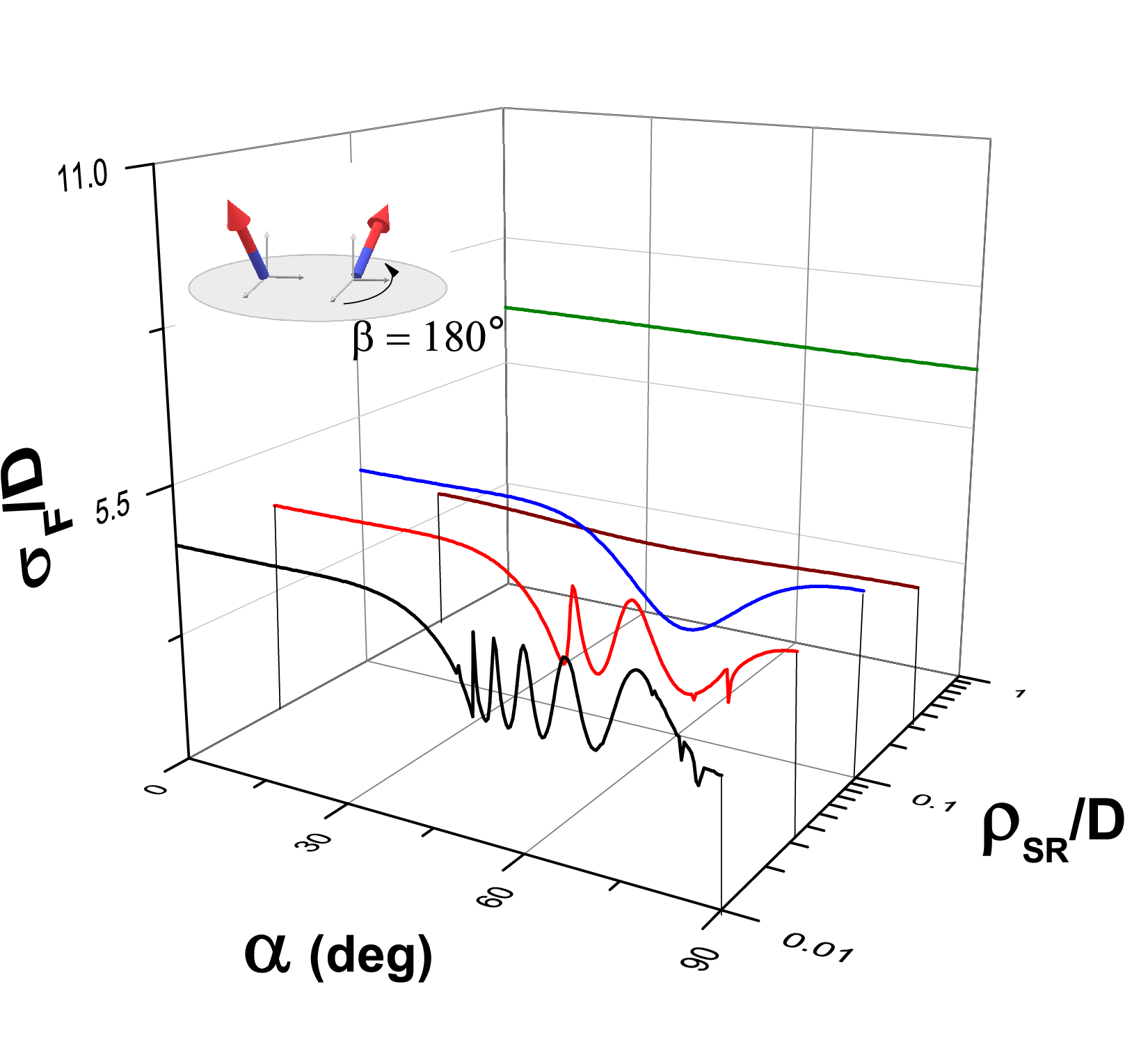}
\put(76,67){\Large{\textbf{(\textit{f})}}}
\end{overpic}
\end{minipage}
\hfill
\vfill
\caption{
The dependencies of the total cross section on the tilt angle $\alpha$ of the dipoles (with respect to the normal to the scattering plane $XY$) for the case of the equal dipole tilt angles $\gamma = \alpha$ at various values of $\rho_{SR}$ in the 2D scattering of identical \textbf{bosons} ($\beta=0^{\circ} (a),90^{\circ} (c),180^{\circ} (e)$) and \textbf{fermions} ($\beta=0^{\circ} (b),90^{\circ} (d),180^{\circ} (f)$). The critical angle $\alpha_c(\beta,\gamma) = 35.3^{\circ},39.2^{\circ},45^{\circ}$ for $\beta = 0^{\circ},90^{\circ},180^{\circ}$, respectively, and $\gamma=\alpha$.
} 
\label{fig3DFullCrossOverAlpha}
\end{figure*}

We also analyzed the SRI impact on the total cross section for the high energies of the dipolar collisions in a plane ($E = 50 E_D$), that were studied in Refs.~\cite{Ticknor_2011,Koval_2014} at fixed values of $\rho_{SR}$. The calculated dependencies of the total cross section on the dipole tilt angle $\alpha$ (at $\gamma=\alpha$), for \textit{variable $\rho_{SR}$ and rotation angle $\beta$} are presented in Fig.~\ref{fig3DFullCrossOverAlpha} for the case of identical bosons and fermions. 
The graph analysis demonstrates the existence of ``threshold'' value $\rho_{SR}/D = 0.3$, below which the strong oscillations (with resonances number increase) occur in the dependencies $\sigma_{B}(\alpha)$, $\sigma_{F}(\alpha)$. 
The number and magnitude of the $\sigma_{B}(\alpha)$, $\sigma_{F}(\alpha)$ oscillations decrease with an increase of the rotation angle $\beta \to 180^{\circ}$. Note, that oscillations and resonances in the dependencies $\sigma_{B}(\alpha)$, $\sigma_{F}(\alpha)$ disappear at the tilt angle $\alpha$ less than the critical angle $\alpha_c(\beta,\gamma)$ at $\gamma=\alpha$ (\ref{eqCriticalAngleForArbitraryBeta}) and $\rho_{SR}/D < 0.1$. That is, in the considered range $\rho_{SR}/D < 1$ \textit{the SRI effects dipolar scattering if the tilt angle $\alpha$ is greater than the critical angle}. The DDI could be seen as a perturbation of the isotropic repulsive part at $\rho_{SR}/D \gg 1$, and this regime will be studied in future works. 
The SRI radius $\rho_{SR}$ decreases with the tilt angle $\alpha$ larger than the critical angle $\alpha_c$ which leads to the occurrence of oscillations and resonances in the dipolar scattering cross section of bosons as well as fermions.
Occurrence of the oscillations and resonances at the angle $\alpha$ achieving the critical angle $\alpha_c$ is demonstrated in Fig.~\ref{fig3DFullCrossOverAlpha} for the range $\rho_{SR}/D \leq 0.1$, when both dipolar interaction and SRI contribute to the scattering (in Fig.~\ref{fig3DFullCrossOverAlpha}(a,b) $\alpha_c = 35.3^{\circ}$, in Fig.~\ref{fig3DFullCrossOverAlpha}(c,d) $\alpha_c = 39.2^{\circ}$, in Fig.~\ref{fig3DFullCrossOverAlpha}(e,f) $\alpha_c = 45^{\circ}$).

\subsection{Angular distributions of differential cross section}
\label{sec:Results:AngularDistributions}

\begin{figure*}[ht]\centering
\hfill
\begin{minipage}[h]{0.45\linewidth}
\begin{overpic}[height=5cm,width=\linewidth,keepaspectratio]
{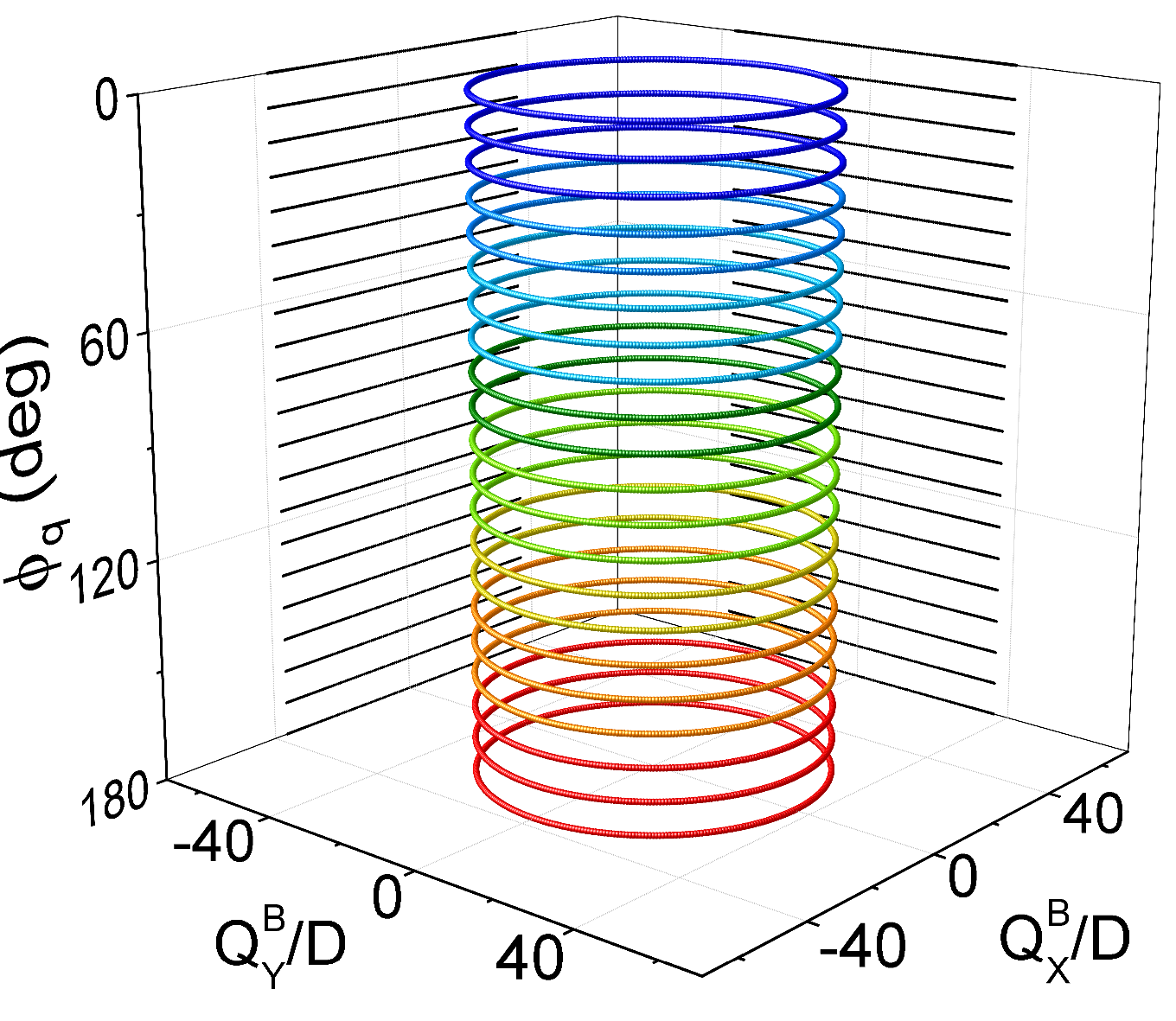}
\put(85,70){\Large{\textbf{(\textit{a})}}}
\end{overpic}
\end{minipage}
\hfill
\begin{minipage}[h]{0.45\linewidth}
\begin{overpic}[height=5cm]{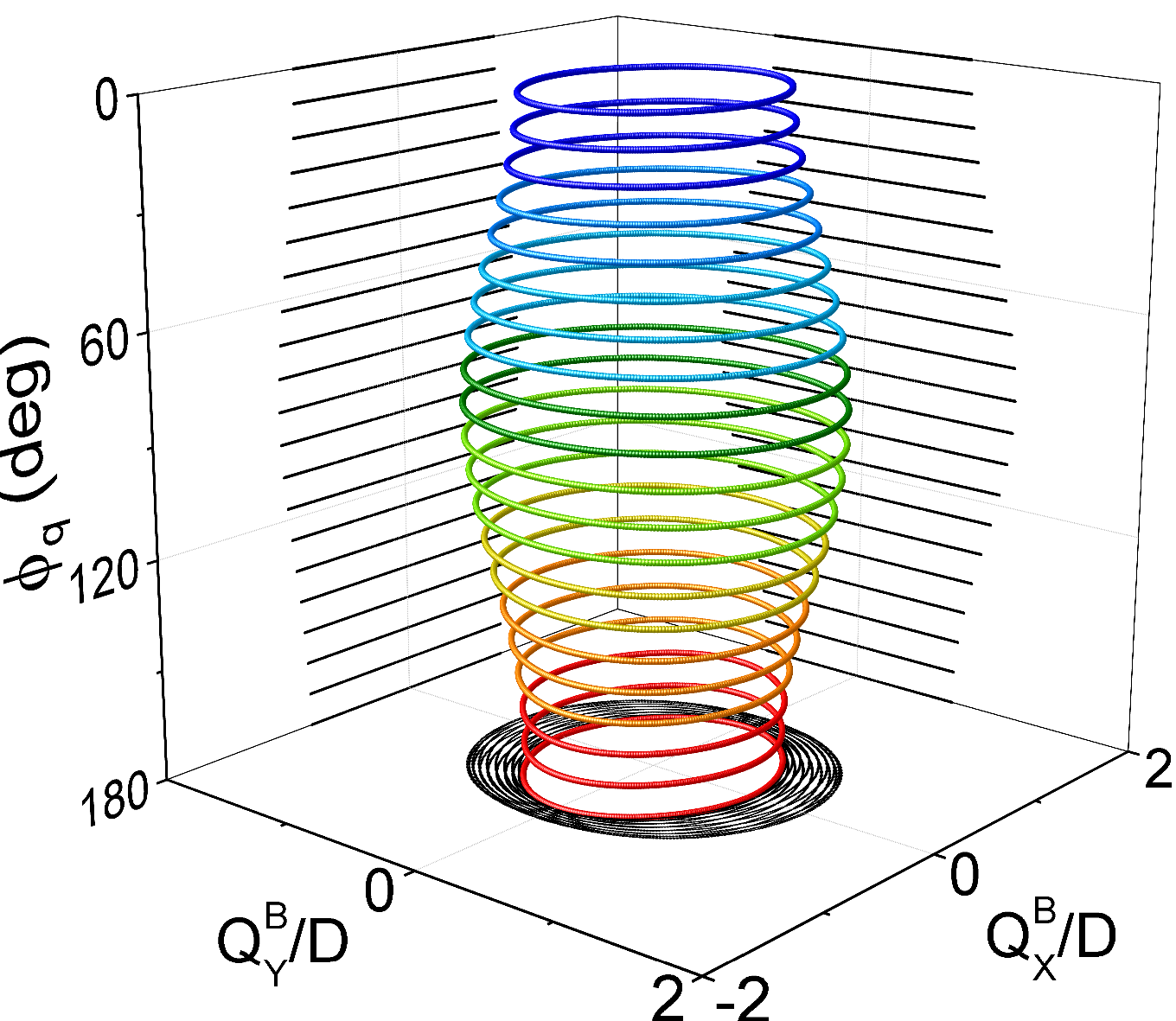}
\put(85,70){\Large{\textbf{(\textit{b})}}}
\end{overpic}
\end{minipage}
\hfill
\vfill
\hfill
\begin{minipage}[h]{0.45\linewidth}
\begin{overpic}[height=5cm,width=\linewidth,keepaspectratio]
{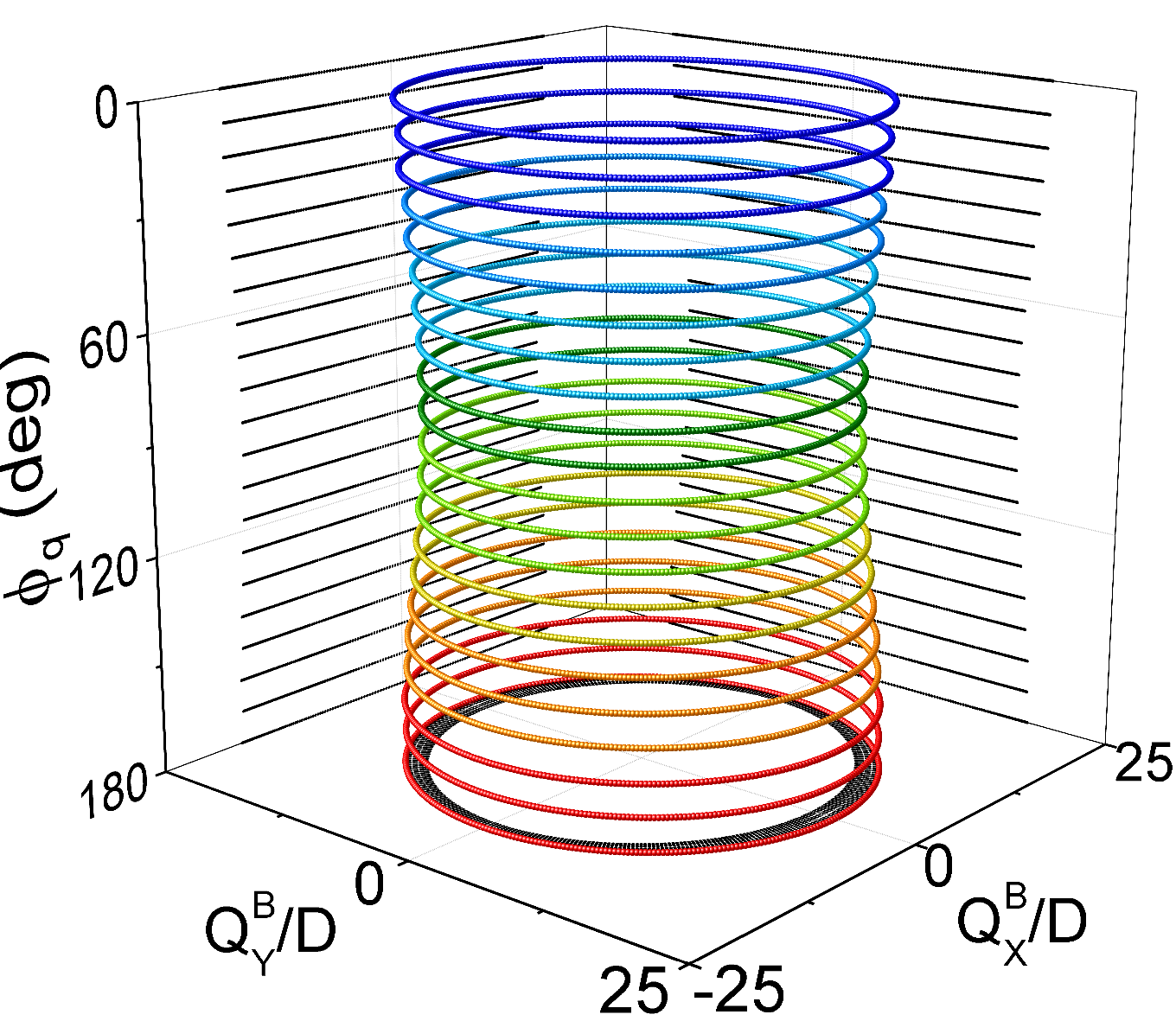}
\put(85,70){\Large{\textbf{(\textit{c})}}}
\end{overpic}
\end{minipage}
\hfill
\begin{minipage}[h]{0.45\linewidth}
\begin{overpic}[height=5cm,width=\linewidth,keepaspectratio]
{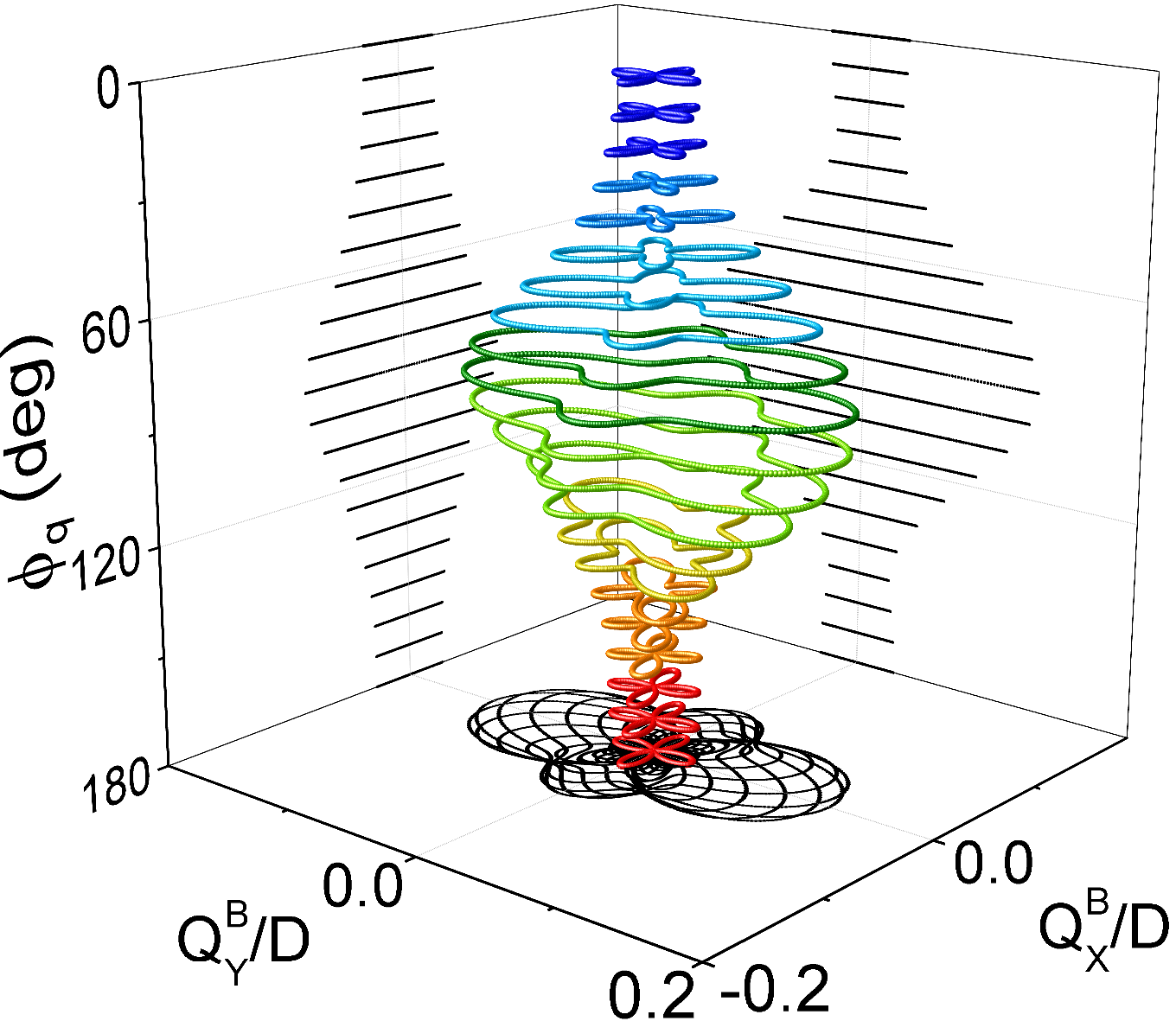}
\put(85,70){\Large{\textbf{(\textit{d})}}}
\end{overpic}
\end{minipage}
\hfill
\vfill
\hfill
\begin{minipage}[h]{0.45\linewidth}
\begin{overpic}[height=5cm,width=\linewidth,keepaspectratio]
{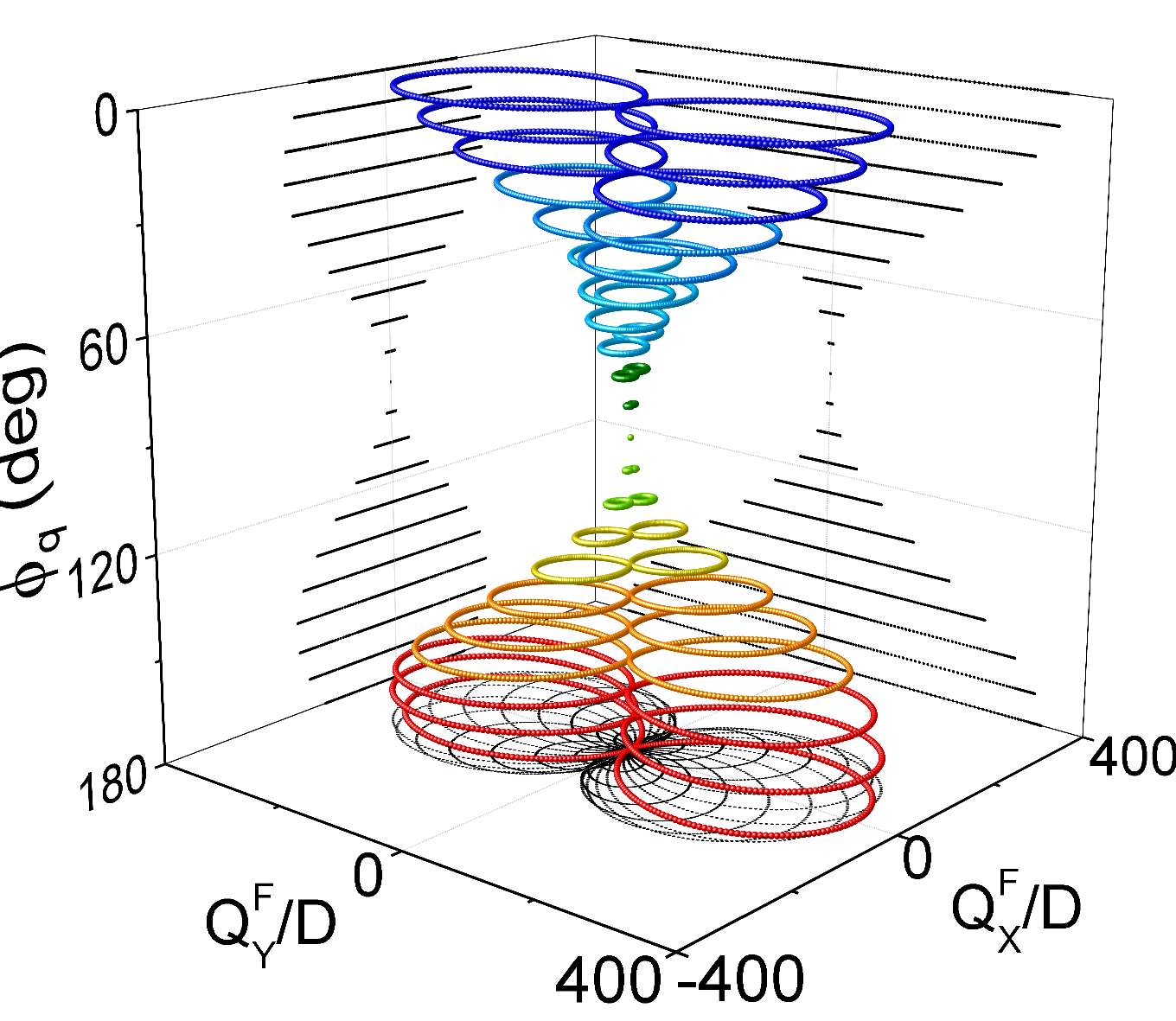}
\put(85,70){\Large{\textbf{(\textit{e})}}}
\end{overpic}
\end{minipage}
\hfill
\begin{minipage}[h]{0.45\linewidth}
\begin{overpic}[height=5cm,width=\linewidth,keepaspectratio]{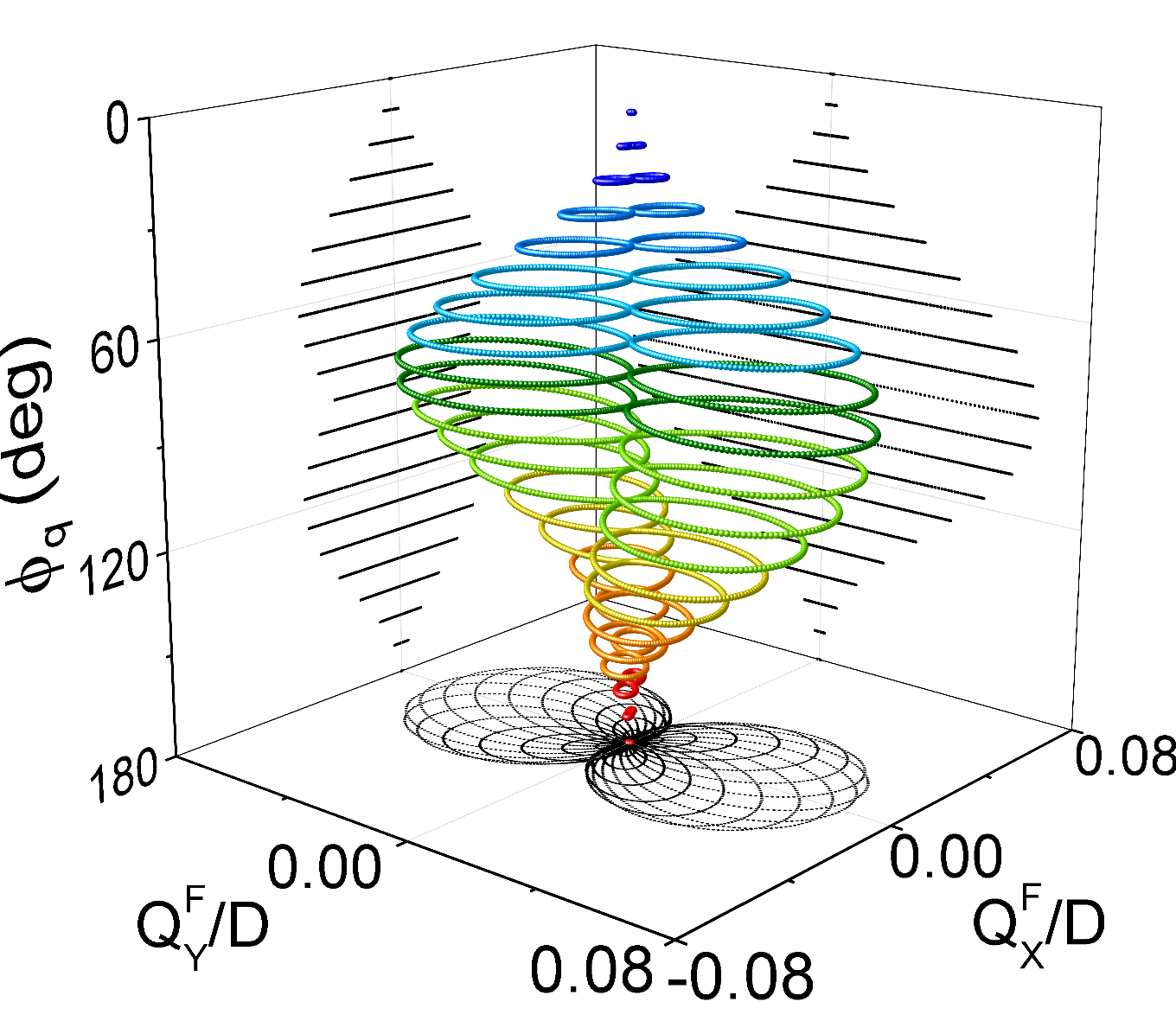}
\put(85,70){\Large{\textbf{(\textit{f})}}}
\end{overpic}
\end{minipage}
\hfill
\vfill
\hfill
\begin{minipage}[h]{0.45\linewidth}
\begin{overpic}[height=5cm,width=\linewidth,keepaspectratio]
{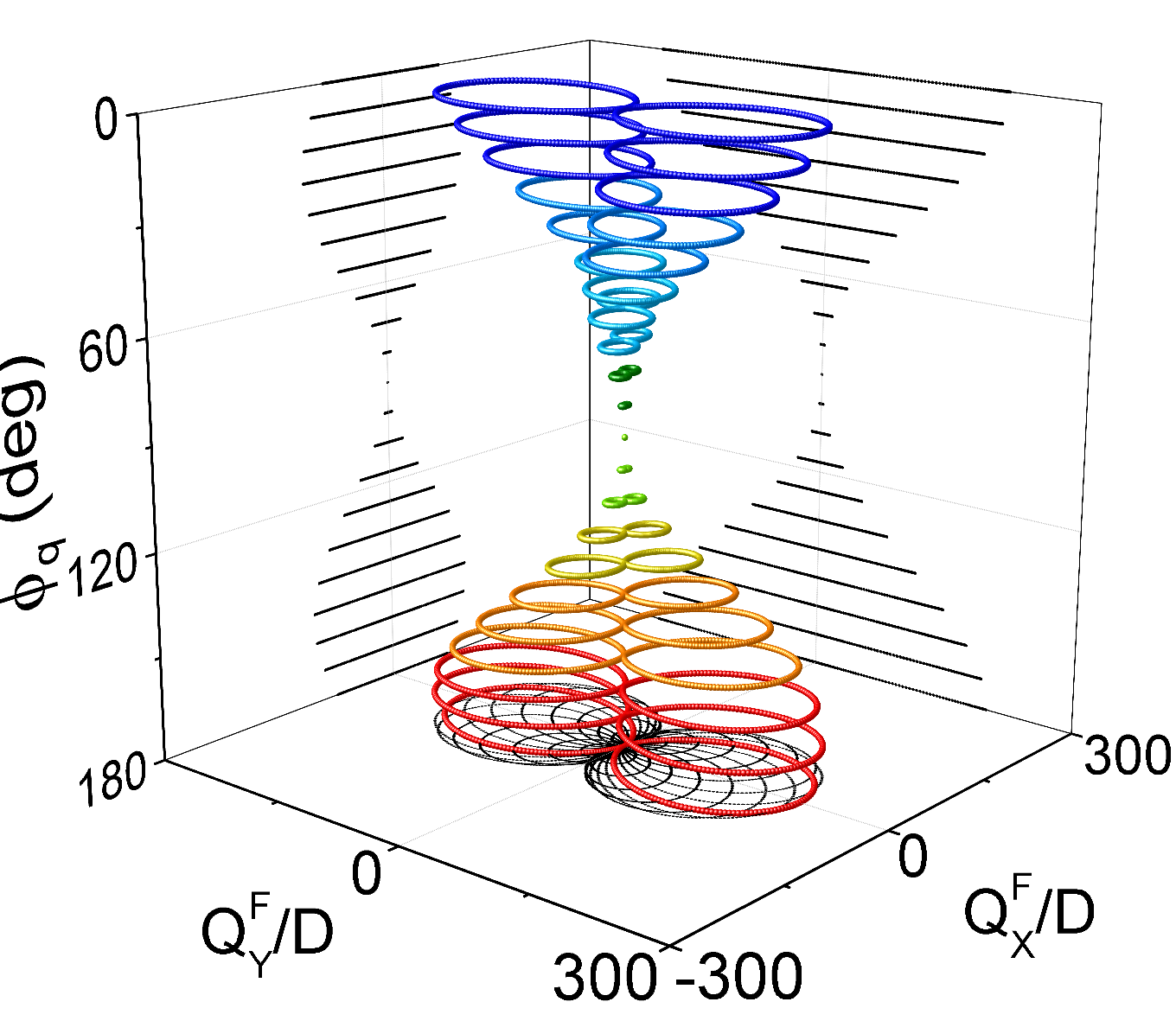}
\put(85,70){\Large{\textbf{(\textit{g})}}}
\end{overpic}
\end{minipage}
\hfill
\begin{minipage}[h]{0.45\linewidth}
\begin{overpic}[height=5cm,width=\linewidth,keepaspectratio]
{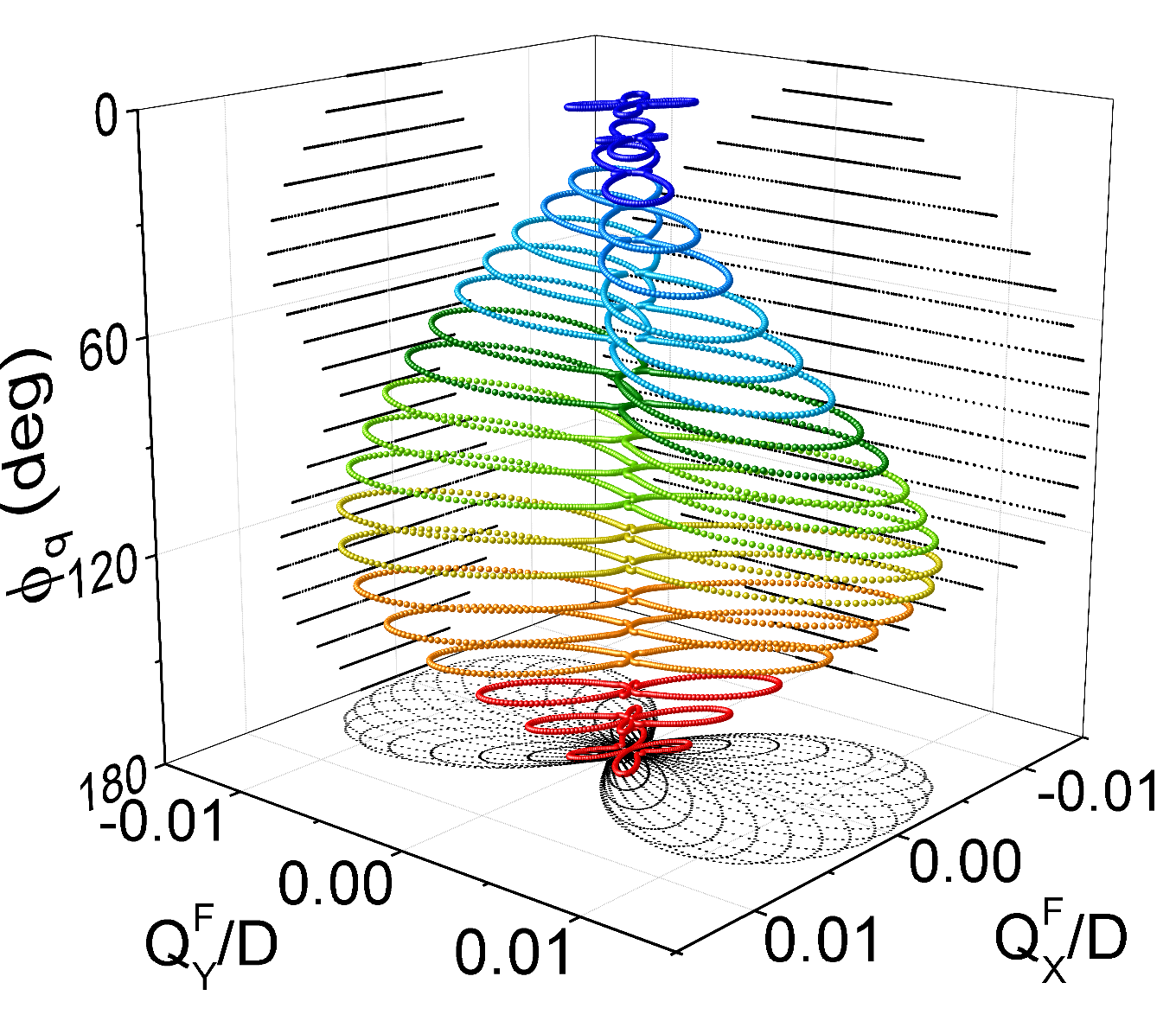}
\put(85,70){\Large{\textbf{(\textit{h})}}}
\end{overpic}
\end{minipage}
\hfill
\vfill 
\caption{The dependencies of the differential cross sections $d\sigma/d\Omega$ on the scattering angle $\phi$ at the different values of the incident angle $\phi_q$, which is changing along the $Z$-axis, for the resonant ($a,c$) and non-resonant ($b,d$) dipolar scattering of bosons; as for resonant ($e,g$) and non-resonant ($f,h$) dipolar scattering of fermions. Here $Q_X^B = d\sigma_B/d\Omega \cos(\phi)$, $Q_Y^B = d\sigma_B/d\Omega \sin(\phi)$ and $Q_X^F = d\sigma_F/d\Omega \cos(\phi)$, $Q_Y^F = d\sigma_F/d\Omega \sin(\phi)$. The data on the graphs is presented as layers for illustrative purposes. The curves are presented for the tilt angles $\alpha=45^{\circ}$ ($a,b,e,f$) and $90^{\circ}$ ($c,d,g,h$) of two aligned ($\beta=0^{\circ}; \gamma = \alpha$) dipoles.
}
\label{fig3DDiffCrossSection}
\end{figure*}

We revealed the strong dependence of the angular distributions of the 2D dipolar scattering differential cross section $d\sigma/d\Omega$ on the value of SRI radius $\rho_{SR}$. 
The differential cross sections of the dipolar scattering of bosons for resonant case 
are presented in Fig.~\ref{fig3DDiffCrossSection}(a,c) at the points $\rho_{SR}/D=0.076, 0.056$ and for non-resonant case are presented
in Fig.~\ref{fig3DDiffCrossSection}(b,d) at the points $\rho_{SR}/D=0.077, 0.23$. In Fig.~\ref{fig3DDiffCrossSection}(a,b,c,d) $Q_X^B$ and $Q_Y^B$ denote $d\sigma_B/d\Omega \cos(\phi)$ and  $d\sigma_B/d\Omega \sin(\phi)$ respectively.

The differential cross section angular distributions for bosons exhibit circular shape in the resonant $\rho_{SR}$ points both for $\alpha=45^{\circ}$ and $\alpha=90^{\circ}$, indicating $s-$wave dominance 
in the resonance emergence. At dipole tilt angles, which are larger than a critical angle, the $d\sigma/d\Omega$ angular distribution has disturbed resonant-like form at the points of $\sigma(\rho_{SR})$ minimum, that is demonstrated in Fig.~\ref{fig3DDiffCrossSection}(b) for $\alpha=45^{\circ}$. Whereas at the tilt angle $\alpha=90^{\circ}$ angular distributions of $d\sigma/d\Omega$ are strongly anisotropic at the points of a minimum of total cross section dependence $\sigma_{B}(\rho_{SR})$, indicating that the $s-$wave contribution is suppressed and the scattering is governed by higher partial waves. So, in contrast to the central potentials, the 2D low-energy dipolar scattering of bosons is strongly anisotropic and its properties are highly sensitive to the SRI radius as well as dipoles mutual orientation.

The angular distributions of differential cross section $d\sigma/d\Omega$ of the dipolar scattering of fermions are always anisotropic. As shown in Fig.~\ref{fig3DDiffCrossSection}(e,g) at the points $\rho_{SR}/D=0.073, 0.415$, respectively, their shapes are almost the same for $\alpha=45^{\circ}$(e) and $90^{\circ}$(g) at the positions of SRI-resonances; while at non-resonant $\rho_{SR}$ points the differential cross section $d\sigma/d\Omega$ changes with increasing $\alpha$, as illustrated in Fig.~\ref{fig3DDiffCrossSection}(f,h) at the points $\rho_{SR}/D=0.25, 0.39$ respectively. In Fig.~\ref{fig3DDiffCrossSection}(e,f,g,h) $Q_X^F$ and $Q_Y^F$ denote $d\sigma_F/d\Omega \cos(\phi)$ and  $d\sigma_F/d\Omega \sin(\phi)$ respectively.
The angular distributions of the 2D dipolar scattering differ significantly from the angular distributions of differential cross sections of the 3D dipolar scattering~\cite{bohn2014differential}. Dipolar fermions can scatter more strongly than dipolar bosons in the 3D case~\cite{bohn2014differential}, whereas in a 2D case the cross section of dipolar scattering of fermions is several orders of magnitude less than the scattering cross section of bosons at low energies. 
The dependencies of the differential cross section of the boson and fermion 2D scattering on the incident angle $\phi_q$ also differ. 
In contrast to the 3D dipolar scattering~\cite{bohn2014differential}, the scattering of bosons in a plane does not depend on the incident angle $\phi_q$ at the resonant points, while it could be highly sensitive to the incident angle in non-resonant points. The 2D scattering of fermions always depends on the incident angle $\phi_q$.

\section{Conclusion}
\label{sec:Conclusions}

The impact of the short-range interaction on the resonances occurrence in the anisotropic dipolar scattering in a plane was numerically investigated \textit{for different orientations of the dipoles} and for a wide range of collision energies. 
We revealed the strong dependence of the cross section on the radius of short-range interaction, which is modeled by a hard wall potential and by the more realistic Lennard-Jones potential. Both potentials showed an almost equal applicability. The analysis of the obtained results showed, that the short-range potential replacement does not change of the resonance structure, leading only to the slight shifts of resonance positions. The results of the accurate numerical calculations of the cross section agree well with the results obtained within the Born and eikonal approximations. 

It was found, that the $s-$wave ($p-$wave) dominates in the angular distributions of the differential cross section at resonance points in the 2D dipolar scattering of identical bosons (identical fermions), whereas the higher partial waves dominate at non-resonant points and the differential cross sections are highly anisotropic.
The cross sections for the fermion scattering in a plane are substantially smaller, than those for bosons at low energies of collisions, like in the low-energy 3D dipolar scattering~\cite{Cavagnero_2009}. At high energies cross sections of a 2D dipolar scattering of fermions are comparable with the cross sections of bosons.

For both low and high collision energies, we reveal the ``threshold'' values of the radius of the short-range interaction, below which there has been an appearance and a growth of the number of resonances. We also defined the critical (magic) tilt angle of one of the dipoles, depending on the direction of the second dipole for arbitrarily oriented dipoles. It was found that resonances arise only when this angle is exceeded. 
In the absence of the resonances the energy dependencies of the boson (fermion) dipolar scattering cross section grows (is reduced) with energy decrease in 2D case, in contrast to the 3D case, where it has the form of a plateau for both bosons and fermions~\cite{Cavagnero_2009}. 
We also showed that the mutual orientation of dipoles strongly affects the form of the energy dependencies. Thus, at an increase of the dipoles tilt angle $\alpha \to 90^{\circ}$ the resonant and non-resonant cross section energy dependencies begin to oscillate, unlike the 3D dipolar scattering~\cite{Cavagnero_2009,bohn2009quasi}, where there is no such oscillations.

Obtained data on the 2D scattering cross sections of arbitrarily oriented dipoles allow us to conclude that the occurrence and the resonances' number \textit{could be controlled} also by varying the radius of the short-range interaction and  the dipoles mutual orientation.

\begin{acknowledgments}
The authors acknowledge the support of the Russian Foundation for Basic Research, Grant No. 19-32-80003. 
\end{acknowledgments}

\end{document}